\documentclass[smallcondensed]{svjour3}    

\smartqed  

\usepackage{graphicx}
\usepackage{color}
\usepackage{natbib}
\begin{document}

\title{Tides and angular momentum redistribution inside low-mass stars hosting planets: a first dynamical model}


\author{A.~F.~Lanza        \and
        S.~Mathis 
}


\institute{A.~F.~Lanza \at
              INAF-Osservatorio Astrofisico di Catania, \\ 
              Via S.~Sofia, 78 - 95123 Catania, Italy \\
              \email{nuccio.lanza@oact.inaf.it} \and
              S.~Mathis \at 
              Laboratoire AIM Paris-Saclay, \\
              CEA/DRF Ð CNRS Ð Universit\'e Paris Diderot, IRFU/SAp Centre de Saclay,\\
	      91191 Gif-sur-Yvette Cedex, France\\
              \email{stephane.mathis@cea.fr}}
\maketitle

\begin{abstract}
We introduce a general mathematical framework to model the internal transport of angular momentum in a star {hosting} a close-in {planetary/stellar} companion. By assuming that the tidal and rotational distortions are small and that the {deposit/extraction of} angular momentum {induced by stellar winds and tidal torques are} redistributed solely by an effective eddy-viscosity { that depends on the radial coordinate}, we can formulate the model in a completely analytic way. {It allows us to compute simultaneously the evolution of the orbit of the companion and of the spin and the radial differential rotation of the star.} An illustrative application to the case of an F-type main-sequence star {hosting} a hot Jupiter is presented. The general relevance of our model to test more sophisticated numerical {dynamical} models and to study the internal rotation profile of exoplanet hosts, submitted to the combined effects of tides and stellar winds, by means of asteroseismology are discussed.\\
~\\


\end{abstract}

\section{Introduction}
\label{intro}
Tides in close binary systems produce a redistribution of  angular momentum and a dissipation of kinetic energy until a final equilibrium state is attained or the two bodies collide with each other \citep[e.g.][]{Hut1980,Hut1981}. The equilibrium state is the minimum kinetic energy state corresponding to the given total angular momentum of the system. It is characterized by the synchronous rotation of the two bodies with their orbit, the alignment of their spin angular momenta with the orbital angular momentum, and a circular orbit. This state is typical of binary systems consisting of two stars of comparable masses \citep[e.g.][]{Zahn1977}. On the other hand, if the initial orbital angular momentum is smaller than three times the sum of the spin angular momenta of the two bodies, the synchronization of their rotation with the orbit is not possible. In this case, the system is unstable \citep{Hut1980} and the tidal evolution will lead to the fall of the less massive body onto the heavier one, provided that the initial rotation period of the latter is longer than the orbital period \citep[e.g.][]{Levrardetal2009}. { In binary systems containing late-type stars, stellar winds will spin down their components, so that tidal equilibrium is not possible and the final fate of the system depends on the wind braking timescale compared to the main-sequence lifetime of the stars. For a discussion of the role of wind braking see, e.g. \citet{Dobbs-Dixonetal04}, \citet{BO2009} and \citet{DL2015}.}

The timescale for attaining the final state (tidal equilibrium or collision) depends on the {physical} processes that dissipate the kinetic energy of the system. In late-type stars and planets with an outer convective envelope, turbulent convection acts on the equilibrium tidal flow, i.e., the flow induced by the {hydrostatic elongation of the body in the direction of the companion, which is steady when rotation is uniform in a reference frame rotating with this latter \citep[e.g.][]{Zahn1966a,RMZ2012} that} produces a net dissipation of its kinetic energy {\citep[e.g.][]{Zahn1989,PS2011,OgilvieLesur2012}}. On the other hand, the time-varying tidal potential, as seen in the same reference frame, excites different kinds of waves ({the so-called dynamical} tide) that {also} contribute to the dissipation of the kinetic energy. {Indeed, both in convective and radiative regions, the equilibrium tide is not solution of the hydrodynamics equations \citep[e.g.][]{Zahn1975,OL2004,Ogilvie2013} and it must be completed by the so-called dynamical tide. In convective regions, it is constituted of inertial waves, which are driven by the Coriolis acceleration, that are excited when the tidal frequency $\hat{\omega} \in [-2\Omega, 2\Omega]$, where $\Omega$ is the rotation frequency of the star \citep{OL2007}. Therefore, an additional dissipation must be taken into account because of the turbulent friction applied by convective eddies on tidal inertial waves. In stellar radiative  zones, the dynamical tide is constituted by internal gravity waves which are influenced by the Coriolis acceleration \citep{Zahn1975,Terquemetal1998,OL2007,Ivanovetal2013,ADMLP2015}. Their dissipation can compete with those of the equilibrium tide and inertial waves in convective regions \citep[e.g.][]{GD1998,OL2007}, especially when massive companions excite high-amplitude waves that may non-linearly break at the center of K- and G-type stars \citep{BO2010,Barker2011,Guillotetal2014}. In low-mass stars, recent works \citep{OL2007,BO2009,Mathisetal2014,Mathis2015a} demonstrated that the dissipation of the dynamical tide vary over several orders of magnitude with the tidal frequency, stellar mass, age and the related internal structure, rotation, and the adopted subgrid modeling for turbulent friction in convective regions.} Therefore, our current estimates {of tidal dissipation in stars} are uncertain by at least 2-3 orders of magnitude {\citep[see e.g. the detailed discussion in][]{Ogilvie2014}}. 

The discovery of close binary systems consisting of a main-sequence star and a close-in massive planet (hot Jupiter) or brown dwarf made it possible to study tidal dissipation processes in a regime of mass ratio previously not accessible to the observations \citep[e.g.][]{MQ1995,Henryetal2000,Charbonneauetal2000,Fabryckyetal2014}. While binary systems consisting of two late-type stars reach their equilibrium state in a timescale much shorter than the main-sequence lifetime of the two stars \citep[][and references therein]{Zahn2008}, hot Jupiter and brown dwarf systems can be often observed far from their final state with the star rotating slower than the orbit, thus allowing us to study their tidal evolution in action \citep[e.g.][]{Bolmontetal2012,DL2015,FMetal2015}. 

To date, tidal evolution studies have been based on the measurement of the rotation in the stellar photospheres, of the orbital period and eccentricity of binary systems, {and of their spin-orbit misalignment \citep[e.g.][]{Winnetal2010, Albrechtetal2012,McQuillanetal2013,McQuillanetal2014,Gizonetal2013,Garciaetal2014,Ceillieretal2016}. However, recent progress in asteroseismic techniques now allow us to begin the investigation of the internal differential rotation in late-type stars \citep[e.g.][]{Benomaretal2015}. On the one hand, the rapidly rotating stars in close stellar binaries are magnetically active, thus hampering the detection of p-mode oscillation \citep{Chaplinetal2011}. On the other hand, stars accompanied by hot Jupiters and brown dwarfs are generally slow rotators thanks to the relatively lower tidal torques that have not been capable to counteract the loss of angular momentum produced by their magnetized stellar winds \citep{DL2015}. Therefore, those stars are {the best} candidates to investigate internal differential rotation and should allow us a measurement of tidal effects on their internal angular momentum distribution by comparing their behavior with that of single stars with similar parameters. In this framework and to fully exploit the possibilities offered by present and future asteroseismic measurements of internal stellar rotation \citep[see e.g.][and references therein]{Raueretal2014}, we need models to predict the effects of tides on the internal angular momentum distribution in late-type stars. {Only very few works have examined the problem. First, \cite{Zahn1966c} has derived an estimation of the possible radial shear at the convective-radiative interface assuming two different solid-body rotations both in the convective envelope and in the radiative core. Next, \cite{GN1989} have  described the progressive synchronization of stellar interiors by tidal internal gravity waves but in the case of early-type stars \citep[see also][]{TK1998}. Finally, more recent works have provided first quantifications of mean zonal flows induced by tidal inertial waves in convective shells \citep{Tilgner2007, Morizeetal2010,Favieretal2014}. 

In the present work, we {propose a first} analytical model to study the effect of tides on the rotation of the radiative {core} and the outer convection zone of a late-type star, assuming that the angular momentum is redistributed inside those zones {in a diffusive way} by {an effective viscosity} depending only on the radial coordinate. {If this model constitutes a simplification of the real picture, where advection by mean meridional flows \citep{Zahn1992,MaederZahn1998,MathisZahn2004} and transport by Reynolds (and Maxwell) stresses induced by tidal waves \citep[e.g.][]{TK1998,Favieretal2014} must be taken into account, it will allow us to unravel the competition between applied tidal and stellar wind torques and provide first estimates of tidally-induced radial gradients of angular velocity, which can be potentially probed with asteroseismology.} In addition to provide potentially new constraints on tidal dissipation and evolution, models of the internal rotation in stars {hosting companions} are important because {tidally-driven} differential rotation {can} produce an additional internal mixing that {may} affect the abundance of light elements such as Lithium, Boron, and Berilllium \citep[e.g.][]{Zahn1994,Brunetal1999,CT2005} and nucleosynthesis in general \citep{Songetal2013,Songetal2016}. Moreover, it {would} play a fundamental role in the amplification of magnetic fields through hydromagnetic dynamo action \citep[e.g.][]{Browningetal2006}.  Last but not least, the analytical formulation of our model makes it useful to test in a near future more complete and complex numerical models of the same processes, thus allowing their validation.\\

In sec. \ref{model}, we first present the hypotheses and assumptions of our physical model and the corresponding mathematical formalism. In sec. 3, we apply our model to the rotational evolution of an F-star hosting a close massive hot-Jupiter and to its simultaneous orbital evolution. Finally, in sec. 4, we present the conclusions and the perspectives of this work.

\section{Physical model}
\label{model}

\subsection{Basic assumptions and hypotheses}
\label{assumptions}
We consider a binary system consisting of bodies of mass $M$ and $m$, respectively. The more massive (primary) body $A$ has mass $M$ and radius $R$, while the least massive $B$ is assumed to be point-like. {As a first step and to simplify the problem,} the relative orbit is assumed to be circular with a semi-major axis $a$ and in the equatorial plane of the primary body (see Fig. \ref{setup}). This is usually the case for stars hosting hot Jupiters because tidal dissipation inside the planets circularizes the orbit on timescales shorter than the main-sequence lifetimes of the systems and the measurements of their projected obliquity based on the Rossiter-McLaughlin effect indicate alignment between the stellar spin and the orbital angular momentum in a substantial fraction of the systems {when $T_{\rm eff}\le 6250\; {\rm K}$, where $T_{\rm eff}$ is the effective temperature of the star} (e.g., \cite{Winnetal2010} and \cite{Albrechtetal2012}).\\ 

\begin{figure}
\centerline{
 \includegraphics[width=8cm,height=5.5cm,angle=0]{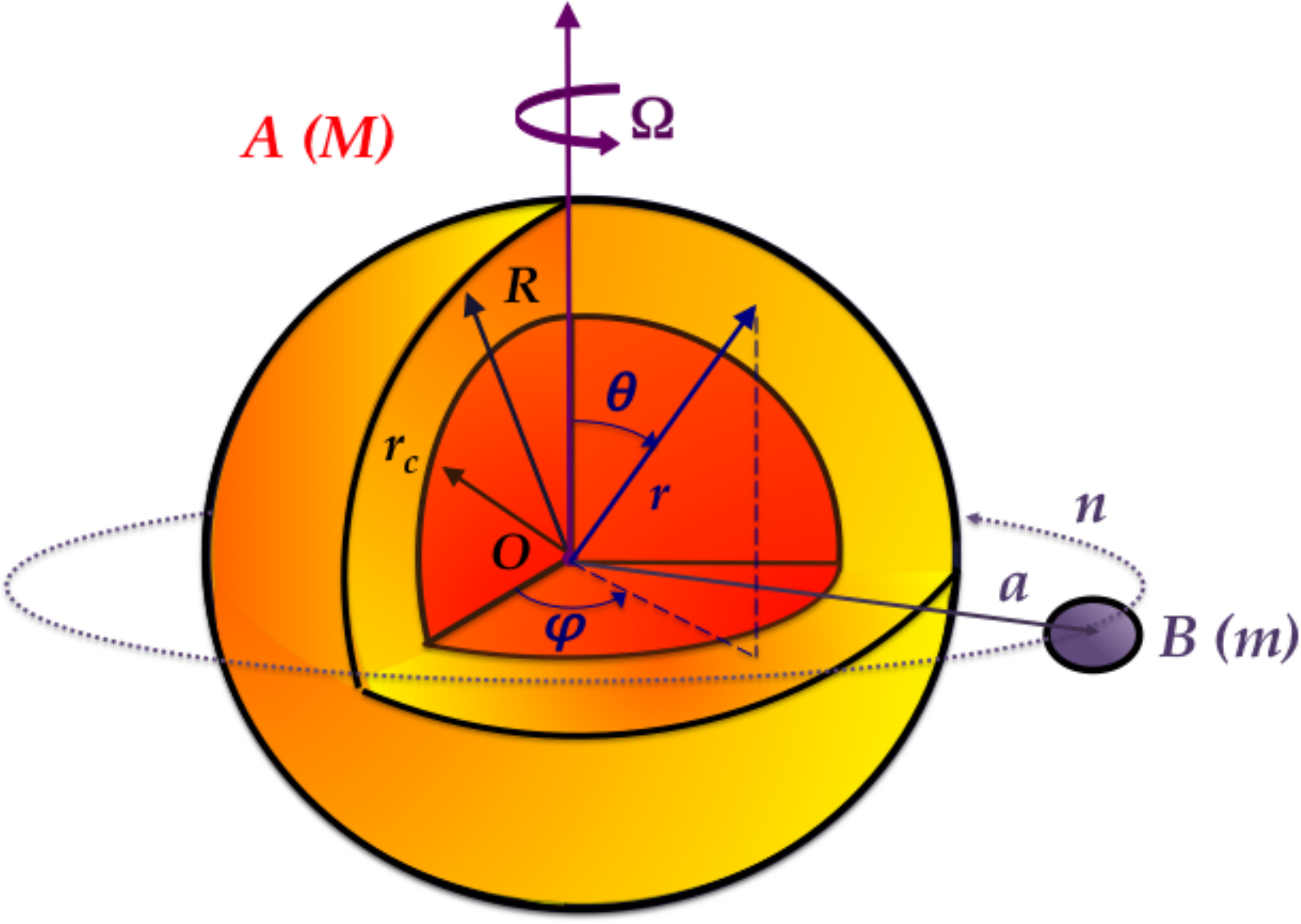}}
 \caption{The studied system: we consider a low-mass star $A$ of mass $M$ and radius $R$ hosting a point-mass companion $B$ of mass $m$ orbiting $A$ circularly in its equatorial plane. The radius of the base (the top) of the convective (radiative) region is $r_{c}$. $\Omega$ is the angular velocity of the star, $n$ the mean motion of the orbit and $a$ its semi-major axis. Finally, we introduce the usual spherical coordinates $\left(r,\theta,\varphi\right)$.}
 \label{setup}
\end{figure}

We shall investigate the angular momentum distribution inside the primary body considering a reference frame having its origin $O$ in the barycentre of that body and its $z$-axis along its spin angular momentum that is parallel to the orbital angular momentum. We adopt spherical polar coordinates $(r, \theta, \varphi)$, where $r$ is the distance from $O$, $\theta$ the colatitude measured from the North pole, and $\varphi$ the azimuthal angle. We assume that the primary body is spherically symmetric, i.e., we neglect the distortions induced by its rotation and the tidal potential of the secondary body. This is not too bad an approximation for stars hosting hot Jupiters because they usually rotate with periods $P_{\rm rot} \sim 5-30$ days, while the system mass ratio is $m/M \sim 10^{-3}$ and $R/a \sim 0.1$. Therefore, their centrifugal distortion is of the order $\epsilon_{\rm C} = \Omega^{2} R^{3}/(GM) \sim 10^{-4} -10^{-5}$, where $\Omega = 2\pi /P_{\rm rot}$ and $G$ is the gravitation constant, while the tidal distortion is $\epsilon_{\rm T} = \delta R/R \sim  (m/M) (R/a)^{3} \sim 10^{-6} - 10^{-5}$. 

If we assume for the moment that the primary body is isolated, the equation for the conservation of its internal angular momentum  can be written in the mean-field approximation as  
\citep[see][and references therein]{Lanza2006,Lanza2007}:
\begin{equation}
\frac{\partial}{\partial t} (\rho r^{2} \sin^{2} \theta \, \Omega) + \nabla \cdot {\vec {\mathcal F}} = 0,  
\label{angmomeq}
\end{equation}
where $\rho(r)$ is the { hydrostatic background} density, $\Omega (r, \theta, t)$ the angular velocity and ${\vec {\mathcal F}}$ 
the angular momentum flux vector given by:
\begin{eqnarray}
\lefteqn{{\vec {\mathcal F}} = (\rho r^{2} \sin^{2} \theta \, \Omega) {\vec u}_{\rm (m)} +} \nonumber \\
 & & + r \sin \theta \langle \rho {\vec u}^{\prime} u_{\varphi}^{\prime} \rangle  
- \frac{r \sin \theta}{\tilde{\mu}} ({\vec B} B_{\varphi} + \langle {\vec B}^{\prime} B_{\varphi}^{\prime} \rangle),
\end{eqnarray}
where ${\vec u}_{\rm (m)}$ is the meridional circulation, ${\vec u}^{\prime}$ the fluctuating velocity field {induced by waves, instabilities or turbulence}, $\tilde{\mu}$
the magnetic permeability, ${\vec B}$ the mean magnetic field and ${\vec B}^{\prime}$ the fluctuating magnetic
field (also induced by waves, instabilities or turbulence)}; angular brackets indicate the Reynolds {and Maxwell stresses azimuthal} averages defining the mean-field quantities. { The mean-field approximation can be justified in the limit of a separation between the small spatial scales of the fluctuations, over which we compute averages, and the large scales that are those over which the mean quantities vary \citep[e.g.][]{RH2004}. Moreover, we filter out any perturbation in time of the density $\rho$, i.e., we adopt the so-called anelastic approximation. 
In the following, we choose for our simplified proof of concept to not model the complex redistribution of angular momentum by meridional flows and internal magnetic fields {both in the radiative core and in the convective envelope \citep[see e.g.][]{BT2002,MathisZahn2004,BMT2004,MathisZahn2005}. In this context, }helioseismology shows that the radiative zone of the Sun is rotating rigidly \citep[e.g.][]{Schouetal1998,Garciaetal2007}. If internal magnetic field has been invoked as a possible cause for this {rotational behavior \citep[see e.g.][]{GM1998,Spruit1999}}, internal gravity waves and some mild {shear-induced} turbulence also appear to be a viable explanation \citep[e.g.][]{TKZ2002}. We thus assume in our model that redistribution of angular momentum in the radiative region results from these two mechanisms, while convective turbulent Reynolds stresses sustain differential rotation in the convective envelope \citep[e.g.][]{BT2002}. In this framework, turbulence and waves'} Reynolds stresses {are written following the mean-field hydrodynamics formalism introduced in \cite{Rudiger1989} and \cite{RH2004}: 
\begin{equation}
\langle \rho u^{\prime}_{i} u^{\prime}_{j} \rangle = -\eta_{\rm eff}\left(r\right)\left( \frac{\partial u_{i}}{\partial x_{j}} +
\frac{\partial u_{j}}{\partial x_{i}} \right) + \Lambda_{ij},  
\label{eq_turb_visc_lambda}
\end{equation}
where ${\vec u}$ is the mean flow field, $\eta_{\rm eff}\left(r\right)$ is {an effective diffusive eddy-viscosity}, assumed to be a scalar function of 
$r$ only, and $\Lambda_{ij}$ indicates the non-diffusive part of the Reynolds stresses due to the velocity
correlations in the rotating star. { To make our model analytically tractable and as a first approximation to the complete problem, we neglect the impact of rotation and magnetic fields on $\eta_{\rm eff}$ (see \citet{Mathisetal2016} for the action of the Coriolis acceleration) and the potential induced anisotropy of $\eta_{\rm eff}$. For example, \citet{Kitchatinovetal94} provide expressions for the components of the anisotropic turbulent diffusivity tensor in the so-called first-order smoothing approximation that neglects the non-linearity of the  interaction between the turbulence, the Coriolis acceleration and the Lorentz force. Moreover, we neglect the role of the $\Lambda_{ij}$ tensor that is a source of angular momentum redistribution that can affect the differential rotation. }

In the case of a non-interacting star, the conservation of the total angular momentum implies:
\begin{equation}
{\mathcal F}_{r} = 0 \mbox{ for $r= R$.}
\label{bcond}
\end{equation}

The equation for the conservation of the angular momentum can be {finally} recast in the form: 
\begin{eqnarray}
\lefteqn{\frac{\partial \Omega}{\partial t} - \frac{1}{\rho r^{4}} \frac{\partial}{\partial r} 
\left(  r^{4} \eta_{\rm eff} \frac{\partial \Omega}{\partial r} \right) +}   \nonumber \\
 &  & -\frac{\eta_{\rm eff}}{\rho r^{2}}
\frac{1}{(1- \mu^{2})} \frac{\partial}{\partial \mu} \left[ (1 - \mu^{2})^{2} \frac{\partial \Omega}{\partial \mu}\right] = S,  
\label{angvel}
\end{eqnarray} 
where $\mu \equiv \cos \theta$ and the source term $S$ is given by:
\begin{equation}
S = - \frac{\nabla \cdot {\vec \tau}}{\rho r^{2} (1- \mu^{2})}; 
\label{sourceangvel}
\end{equation}
${\vec \tau}$ is a vector whose components are: 
\begin{equation}
\tau_{i} = r \sin \theta \Lambda_{i \varphi} + \left[\rho r^{2} \sin^{2} \theta \Omega  u_{\rm (m) i}\right], 
\label{tau}
\end{equation}
where the second term related to large-scale meridional flows is here neglected as discussed above. 
From a physical point of view, $\tau$ is the net flux of angular momentum coming out from the unit volume in the unit time, while $S$ is the $z$-component of the torque per unit volume divided by the moment of inertia of the same volume in order to give the variation of the angular velocity there. 

We shall consider the variation of the angular momentum inside the star in a shell extending from a base radius $r=r_{\rm b}$ to the surface $r=R$ because it is convenient to avoid singularities at $r=0$ in the {numerical} integration of the equations (see Sect.~\ref{ill_case}). This is {for example }applicable to a main-sequence F-type star that has a convective core, an extended radiative zone, and a thin outer convection zone. Taking $r_{\rm b}$ equal to the radius of its inner convective core, we can assume that its total angular momentum stays constant because all  torques by the stellar wind and the tides are applied to the outer convection zone or to the bulk of the radiative zone with their effects propagating very slowly towards the central region of the star, thus having a negligible effect on the rotation of the inner core {(see the detailed discussion in Sec. \ref{ill_case})}. 
Therefore, the boundary conditions given by Eq.~(\ref{bcond}) can be written as:
\begin{equation}
\frac{\partial \Omega}{\partial r} = 0 \mbox{ for $r=r_{\rm b}, \, R$,}
\label{bcond1}
\end{equation}
that imply the conservation of the total angular momentum in the shell between $r_{\rm b}$ and $R$. 

Rigorously speaking, the surface of the star at $r=R$ is not exactly stress-free because the stellar winds and the tides induce torques there. Nevertheless, it is mathematically convenient to assume stress-free boundary conditions on the angular velocity $\Omega$ and add the effects of the wind and tides as internal torques that act inside the outer convection zone of the star (see below).\\ 

Then, we choose to expand the angular velocity as the superposition of a differential rotation $\Omega_{0}\left(r,\mu,t\right)$ sustained by internal processes redistributing angular momentum and a second component $\omega\left(r,\mu,t\right)$ induced by applied tidal and wind torques: 
\begin{equation}
\Omega (r, \mu, t) = \Omega_{0} (r, \mu,t) + \omega (r, \mu, t).  
\end{equation}
Such an expansion is justified in the following situations. First, in convective regions, the characteristic dynamical time of transport of angular momentum by turbulent Reynolds stresses is much shorter than the secular ones associated to the braking of stars by winds and to their spin-up/spin-down by tides. Then, a timescale separation can be assumed when solving Eq. (\ref{angmomeq}) with one equation for short timescales corresponding to convective motions and a separated one for secular timescales. The same applies to stellar radiation zones if the processes suspected to enforce the observed uniform rotation in the Sun \citep[e.g.][]{Garciaetal2007} and the weak differential rotation in other solar-type stars \citep{Benomaretal2015} have characteristic timescales shorter than those of wind and tidal torques. Such an expansion is also possible in stellar radiative regions where $\omega$ is a small fluctuation compared to the mean sustained weak differential rotation.  Moreover, our separation of $\Omega_{0}$ from $\omega$ is valid only as a first approximation if neglecting the dependance of  the turbulent viscosity on rotation. We also neglect the effects of the rotation  on the  $\Lambda$ tensor in Eq.~(\ref{eq_turb_visc_lambda}) that can induce an additional differential rotation component that varies with the stellar spin during the course of the stellar evolution. Moreover, the intensity of the tidal torque could depend explicitly on the local background rotation because of the tidal excitation of inertial and gravito-inertial waves discussed in Sect.~\ref{intro} and of the modification  of the turbulent convective tidal friction by the Coriolis acceleration \citep[][]{Mathisetal2016}. It will be here modeled by a fixed constant modified tidal quality factor computed independently (see Eq. \ref{simplifiedtides}).\\ 

The equation for $\omega$ then becomes: 
\begin{eqnarray}
\lefteqn{\frac{\partial \omega}{\partial t} - \frac{1}{\rho r^{4}} \frac{\partial}{\partial r} 
\left(  r^{4} \eta_{\rm eff} \frac{\partial \omega}{\partial r} \right) +}   \nonumber \\
 &  & -\frac{\eta_{\rm eff}}{\rho r^{2}}
\frac{1}{(1- \mu^{2})} \frac{\partial}{\partial \mu} \left[ (1 - \mu^{2})^{2} \frac{\partial \omega}{\partial \mu}\right] = S_{1},  
\label{angvelp}
\end{eqnarray} 
where the perturbation of the source term is:
\begin{equation}
S_{1} =  - \frac{\nabla \cdot ({\vec \tau}_{\rm W} + \vec \tau_{\rm T})}{\rho r^{2} (1- \mu^{2})},
\label{s1}
\end{equation}
where $\nabla \cdot \vec \tau_{\rm W}$ and $\nabla \cdot \vec \tau_{\rm T}$ are the torques per unit volume induced by the stellar wind and the tides, respectively.  Eq.~(\ref{angvelp})
must be solved together with the boundary conditions:
\begin{equation}
\frac{\partial \omega}{\partial r} = 0 \mbox{ for $r=r_{b}, \, R$.}
\label{bcond2}
\end{equation}
As a consequence of Eq.~(\ref{bcond2}), if we assume that $\eta_{\rm eff}$ vanishes for $r=r_{\rm b}$ and $r= R$, by deriving both sides of Eq.~(\ref{angvelp}) with respect to $r$, we obtain:
\begin{equation}
\frac{\partial S_{1}}{\partial r} = 0 \; \mbox{for $r=r_{\rm b}, R$. }
\end{equation}

\subsection{Solution of the angular momentum equation }
\label{general_sol}

The general solution of Eq.~(\ref{angvelp}) with the boundary conditions (\ref{bcond2}) can be obtained by the method of
separation of the variables and expressed as a series of the form \citep[e.g.][]{Lanza2006}:
\begin{equation}
\omega (r, \mu , t) = \sum_{n = 0, 2, 4, ...}^{\infty} \sum_{k=0}^{\infty} \alpha_{nk} (t) \zeta_{nk} (r) P_{n}^{(1,1)} (\mu),
\label{omegadevel}
\end{equation}
where $\alpha_{nk}(t)$ and $\zeta_{nk}(r)$ are functions  that will be specified below 
and $P_{n}^{(1, 1)}(\mu)$ are Jacobian polynomials, i.e., the finite solutions of the equation: 
\begin{equation}
\frac{d}{d \mu} \left[ (1- \mu^{2})^{2} \frac{d P_{n}^{(1, 1)}}{d \mu }\right] + 
n(n+3) (1 - \mu^{2}) P_{n}^{(1, 1)} = 0, 
\label{jacobian}
\end{equation} 
in the interval $ -1 \leq \mu \leq 1$ including its ends \citep[e.g.][]{AS1972}. 
The Jacobian polynomials form a complete and orthogonal set in this interval with 
respect to the weight function $(1-\mu^{2})$. Only the polynomials of even degree appear in Eq.~(\ref{omegadevel})
because the angular velocity perturbation is symmetric with respect to the equator under the assumptions of our model. 
For $n \gg 1$ the asymptotic expression of the Jacobian polynomials is \citep{GR1994}:
\begin{equation}
P_{n}^{(1,1)} (\cos \theta) =
 \frac{\cos \left[ (n + \frac{3}{2}) \theta 
- \frac{3\pi}{4}\right]}{\sqrt{\pi n} \left[\sin \left(\frac{\theta}{2}\right) \cos \left(\frac{\theta}{2}\right)\right]^{3/2}} 
+ O( n^{-\frac{3}{2}}). 
\label{jacobi_asymp}
\end{equation}

The functions $\zeta_{nk}$ are the solutions of the Sturm-Liouville 
problem defined in the interval $r_{b} \leq r \leq R$ by the equation:
\begin{equation}
 \frac{d}{dr} \left( r^{4} \eta_{\rm eff} \frac{d \zeta_{nk}}{dr} \right) - n(n+3) r^{2} \eta_{\rm eff}
\zeta_{nk} + \lambda_{nk} \rho r^{4} \zeta_{nk} = 0 
\label{radialpart}
\end{equation}
with the boundary conditions (following from Eq.~\ref{bcond2}): 
\begin{equation}
\frac{d \zeta_{nk}}{dr} = 0 \mbox{ at $r=r_{b}, R$.}
\label{bc1}
\end{equation} 
We shall consider normalized eigenfunctions, i.e.: $\int_{r_{\rm b}}^{R} \rho r^{4} \zeta_{nk}^{2} dr = 1$. 
The eigenfunctions $\zeta_{nk}$ for a fixed $n$, form a complete and orthonormal set 
in the interval $[r_{\rm b}, R]$ with respect to the weight function $\rho r^{4}$ that does not depend on $n$. 
 For $n=0$, the first eigenvalue
corresponding to the eigenfunction $\zeta_{00}$ is zero and the eigenfunction is constant, as it is evident by integrating both sides of Eq. (\ref{radialpart}) in the same interval and applying the 
boundary conditions (\ref{bc1}). We shall see below how this constant is related to the total angular momentum of the star. 

We recall
from the theory of the Sturm-Liouville problem that the eigenvalues for fixed $n$ verify the inequality:
$ \lambda_{n 0} < \lambda_{n 1} <...<\lambda_{n k}< \lambda_{n k+1}< ... $ and that the
eigenfunction $\zeta_{n k}$ with  $k> 0$ has $k-1$ nodes in the interval $[r_{b}, R]$ for each $n$, if we indicate with $\zeta_{00}$ the constant eigenfunction associated with $\lambda_{00}=0$ {as  discussed above}.  For $n >0$, all the eigenvalues $\lambda_{n0}$ are
positive, as can be derived by integrating both sides of Eq. (\ref{radialpart}) in the  interval $[r_{b}, R]$,
applying the boundary conditions (\ref{bc1}) and considering that $\zeta_{n0}$ has no nodes. In view of the inequality
given above, all the eigenvalues $\lambda_{nk}$ are then positive for $ n > 0$. Moreover, it is possible to prove
that  $\lambda_{n^{\prime} k} \geq \lambda_{n k}$ if $n^{\prime} > n$ {\citep[see][]{MF1953}}. 
 For  $k \gg 1$ and $k \gg n$ the asymptotic expression for
the eigenvalue is independent of $n$ and is:
\begin{equation}
\lambda_{nk} \simeq \frac{\pi^{2} k^{2}}{\int_{r_{\rm b}}^{R} \sqrt{\rho/\eta_{\rm eff}} d r^{\prime}},
\label{eigen_asymp}
\end{equation}
while the 
eigenfunction $\zeta_{nk}$ is \citep[e.g.][]{MF1953}:
\begin{equation}
\zeta_{nk}(r) \simeq \left( \rho \eta_{\rm eff}\right)^{-\frac{1}{4}} r^{-2} \cos \left( \sqrt{\lambda_{nk}}
 \int_{r_{b}}^{r} \sqrt{\rho/\eta_{\rm eff}} d r^{\prime} \right). 
\label{zeta_asymp}
\end{equation}  
 
The time dependence of the solution (\ref{omegadevel}) is specified by the functions $\alpha_{nk}$ that are given by:
\begin{equation}
\frac{d \alpha_{nk}}{dt} + \lambda_{nk} \alpha_{nk}(t) = \beta_{nk} (t),
\label{alphaeq}
\end{equation}
where the functions $\beta_{nk}$ appear in the development of the {forcing} term $S_{1}$:
\begin{equation}
S_{1}(r, \mu, t) = \sum_{n} \sum_{k} \beta_{nk} (t) \zeta_{nk} (r) P_{n}^{(1,1)}(\mu),
\label{s1devel}
\end{equation}
and are given by \cite{Lanza2006}:
\begin{eqnarray}
\lefteqn{\beta_{nk} = \frac{(2n+3) (n+2)}{8(n+1)} \, \times} \nonumber \\ 
& & \times \int_{r_{\rm b}}^{R} \int_{-1}^{1} \rho {r'}^{4} (1-{\mu'}^{2}) S_{1}({r'}, {\mu'}, t) \zeta_{nk}({r'}) 
P_{n}^{(1,1)} ({\mu'}) d{r'} d{\mu'}. 
\label{betaeq}
\end{eqnarray}
The solution of Eq.~(\ref{alphaeq}) is:
\begin{equation}
\alpha_{00} (t) = \alpha_{00} (0) + \int_{0}^{t} \beta_{00} (t^{\prime}) d t^{\prime},
\label{alphabeta0}
\end{equation}
for the coefficient associated with the first eigenfunction, whose eigenvalue is zero, while:
\begin{equation}
\alpha_{nk}(t) = \alpha_{nk} (0) \exp(-\lambda_{nk} t) + \exp(-\lambda_{nk} t) \int_{0}^{t} \beta_{nk}(t^{\prime}) \exp(\lambda_{nk} t^{\prime})
 d t^{\prime} 
\label{alphabeta}
\end{equation}
for $\lambda_{nk}> 0$, which allows us to specify the general solution of Eq.~(\ref{angvelp}) with the boundary conditions 
(\ref{bcond2}) when the {forcing} term $S_{1}(r, \mu, t)$ and the 
initial conditions are given. 

If the internal rotation is initially non-uniform and the source term $S_{1}$ vanishes everywhere, the angular momentum redistribution due to the {effective eddy-}viscosity drives the system towards a final state of uniform rotation because the coefficients $\alpha_{nk}$ of all the eigenfunctions with $k>0$ decay exponentially in time (cf. Eq.~\ref{alphabeta} with $\beta_{nk}=0$). The eigenfunction with the slowest decay rate is that with the lowest eigenvalue, i.e., $\lambda_{01}$, that sets the characteristic timescale of angular momentum redistribution across the interval $[r_{\rm b}, R]$, that is $\lambda_{01}^{-1}$. This is the characteristic timescale after which any initial spatial inhomogeneity in the angular velocity is erased by the effective diffusion. 

When the functions $\beta_{nk}(t)$ are non-zero and vary slowly with respect to $\lambda_{nk}^{-1}$, i.e., the characteristic timescale of diffusion of the angular momentum associated with the eigenfunction $\zeta_{nk}$, we can take $\beta_{nk}(t)$ outside of the integration and obtain the approximate solution:
\begin{equation}
\alpha_{nk}(t) \simeq \alpha_{nk} (0) \exp(-\lambda_{nk} t) + \beta_{nk}(t)[1- \exp(-\lambda_{nk} t) ] \lambda_{nk}^{-1}
\label{slow_solution}
\end{equation}
that gives the stationary solution for $t \gg \lambda_{nk}^{-1}$ , independent of the initial rotation profile: 
\begin{equation}
\alpha_{nk} (t) \simeq \beta_{nk}(t) \lambda_{nk}^{-1}.
\label{long_timescale}
\end{equation}

The case when the diffusion timescales associated with the eigenvalues are comparable or shorter than the timescale of variation of $\beta_{nk}$ requires to perform the integration indicated in Eq.~(\ref{alphabeta}). For a variation of the stellar angular momentum produced by a magnetized wind or tides, it is possible to perform the integration analytically, if we adopt the simple parametrizations given in Sect.~\ref{lat_ave_model}. Nevertheless, the primitive functions involve complicated transcendental functions such as the imaginary error function or the incomplete gamma function\footnote{For more details, the reader can apply the integral calculator available through the Wolfram website at: http://www.wolframalpha.com/calculators/integral\_calculator.}. 

\subsection{Variation of the total angular momentum}
The rate of variation of the total angular momentum $L$ of our primary star is:
\begin{equation}
\frac{dL}{dt} = \int_{V} \rho {r'}^{2} (1- {\mu'}^{2}) \frac{\partial \omega}{\partial t} dV,
\end{equation}
where $V$ is its volume. If we substitute the expression for the angular velocity $\omega$ as given by Eq.~(\ref{omegadevel}) and consider that the elementary volume is $d V = -r^{2} d\mu \, dr\, d \varphi$, we obtain:
\begin{equation}
\frac{dL}{dt} = \frac{8\pi}{3} \sum_{k=0}^{\infty} \frac{d \alpha_{0k}}{dt} \int_{r_{\rm b}}^{R} \rho r^{\prime \,4}\zeta_{0k} d r^{\prime}, 
\label{totang0}
\end{equation}
where we have made use of the orthogonality of the Jacobian polynomials with respect to the weight function $1-\mu^{2}$ and considered that $P_{0}^{1,1}(\mu) =1$. 
For $k >0$, all the definite integrals appearing on the r.h.s. of Eq.~(\ref{totang0}) vanish as can be immediately found by integrating Eq.~(\ref{radialpart}) in the interval $[r_{\rm b}, R]$ and applying the boundary conditions (\ref{bc1}). Therefore, only the term associated with the first constant eigenfunction $\zeta_{00}$ remains. 
On the other hand, considering the normalization of $\zeta_{00}(r)$, we have: 
\begin{equation}
\frac{dL}{dt} = \frac{8\pi}{3} \frac{d\alpha_{00}}{dt} \left( \int_{r_{\rm b}}^{R} \rho r^{\prime \, 4} \, dr^{\prime} \right)^{1/2}= \frac{8\pi}{3} \beta_{00}(t) \left( \int_{r_{\rm b}}^{R} \rho r^{\prime \, 4} \, dr^{\prime} \right)^{1/2}. 
\label{tot_ang_mom_var}
\end{equation}

The moment of inertia $I_{\rm s}$ of the shell $[r_{\rm b}, R]$ is given by:
\begin{equation}
I_{\rm s} = \frac{8\pi}{3} \int_{r_{\rm b}}^{R} \rho r^{\prime \, 4}\, dr^{\prime}.
\end{equation}
Introducing the mean angular velocity of the shell as $\omega_{\rm mean} \equiv L/I_{\rm s}$, we finally obtain:
\begin{equation}
\frac{d \omega_{\rm mean}}{dt} = \zeta_{00} \frac{d \alpha_{00}}{dt} = \left( \int_{r_{\rm b}}^{R} \rho r^{\prime \, 4} \, dr^{\prime} \right)^{-1/2} \beta_{00}(t). 
\end{equation}

\subsection{Latitudinally averaged model}
\label{lat_ave_model}
It is interesting to consider a convenient latitudinal average of the equation for the angular velocity because current asteroseismic techniques provide an estimate of the radial gradient of the angular velocity inside stars \citep[see e.g.][and references therein]{Aerts2015}. This happens because only  modes with a degree $\ell \leq 3$ are detectable by means of photometric observations, so that any resolution in latitude is lacking \citep[e.g.][]{Benomaretal2015}.  The simplest latitudinal average can be obtained by multiplying Eq.~(\ref{angvelp}) by $1-\mu^{2}$, that gives the latitudinal dependence of the moment of inertia, and integrating over $\mu$. The angular part of the equation is {thus} averaged out and we obtain:
\begin{equation}
\frac{\partial \tilde{\omega}}{\partial t} -\frac{1}{\rho r^{4}} \frac{\partial}{\partial r} \left( r^{4} \eta_{\rm eff} \frac{\partial \tilde{\omega}}{\partial r} \right) = \tilde{S}_{1},
\label{lataveeq}
\end{equation}
where the latitudinal average of the angular velocity is defined as:
\begin{equation}
\tilde{\omega}\left(r,t\right) \equiv \int_{-1}^{1} \omega(r, {\mu'}, t) (1 -{\mu'}^{2}) d{\mu'}, 
\end{equation}
while the latitudinally averaged torque is:
\begin{equation}
\tilde{S}_{1} \equiv \int_{-1}^{1} S_{1}(r, {\mu'}, t) (1-{\mu'}^{2}) \, d{\mu'}. 
\end{equation}
Eq.~(\ref{lataveeq}) must be solved with the boundary conditions: $\partial \tilde{\omega} / \partial r = 0$ at $r=r_{\rm b}, R$. 

Thanks to the orthogonality properties of the Jacobian polynomials and considering that $P_{0}^{1,1}(\mu) = 1$, we can develop the solution $\tilde{\omega}$ of Eq.~(\ref{lataveeq}) into a series that  
contains only the eigenfunctions $\zeta_{0k}$ as:
\begin{equation}
\tilde{\omega} = \sum_{k=0}^{\infty} \alpha_{0k}(t) \zeta_{0k}(r),
\end{equation}
where the time-dependent coefficients $\alpha_{0k}$ obey  Eqs.~(\ref{alphabeta0}) and~(\ref{alphabeta}); the boundary conditions are automatically satisfied given the definition of the eigenfunctions $\zeta_{0k}$.  

The development of $\tilde{S}_{1}$ into a series of the same eigenfunctions is given by:
\begin{equation}
\tilde{S}_{1} = \sum_{k=0}^{\infty} \beta_{0k} (t) \zeta_{0k} (r),
\end{equation}
where the coefficients $\beta_{0k}$ are given by Eq.~(\ref{betaeq}) with $n=0$ and $P^{(1,1)}_{0}=1$. 

We can derive expressions for the terms $\tilde{S}_{1 \rm W}$ and $\tilde{S}_{1 \rm T}$ produced by the stellar wind and the tides, respectively, by considering the relationship between $\tilde{S}_{1}$ and the variation of the angular momentum $dL$ in a spherical shell between radii $r$ and $r+dr$:
\begin{equation}
\frac{d}{dt} \left( \frac{dL}{dr} \right) = 2 \pi \rho r^{4} \tilde{S}_{1} (r, t),
\label{ave_s1}
\end{equation}
that is obtained from Eq.~(\ref{lataveeq}) by considering that the angular momentum of the shell obeys the relationship: $dL/dr = 2 \pi \rho r^{4} \tilde{\omega}$.\\  

{The first process responsible for the rotation evolution of late-type stars is the angular momentum loss produced by their magnetized winds (see, e.g., \cite{GB2013}, \cite{GB2015}  and \cite{Mattetal2015} for details). We consider a simplified treatment by assuming that, 
after an initial phase, the rotation period of a main-sequence star varies according to the so-called Skumanich law, that is $P_{\rm rot} \propto \sqrt{t}$, where $t$ is the age of the star \citep{Sku1972,Kawaler1988}.} The corresponding angular momentum loss rate is proportional to $\Omega^{3}$, assuming a constant moment of inertia of the star. Turbulent convection redistributes the angular momentum loss across the whole convection zone over a timescale comparable with the convective turnover time, i.e., much shorter than the timescale for the evolution of the stellar angular momentum. Therefore, we can assume that the torque is uniform across a radial interval $[r_{c}, R]$, where $r_{c}$ is the base of the  convection zone. In other word, the angular momentum loss associated with the stellar wind can be expressed as:
\begin{equation}
\frac{d}{dt} \left( \frac{dL_{\rm W}}{dr} \right) = -\frac{K_{\rm wind} \Omega^{3}}{R-r_{c}},
\end{equation}
where $K_{\rm wind}$ is a constant depending on the mass of the star and $\Omega$ its surface angular velocity that is a function of the time. From this equation, we derive the corresponding $\tilde{S}_{1 \rm W}$ term from Eq.~(\ref{ave_s1}).\\ 

Similarly, we can express the tidal contribution to $\tilde{S}_{1}$. The tidal torque {in the coplanar and circular case} in the weak-friction approximation is proportional to $\left(\Omega -n\right)$, {where $n$ is the mean orbital motion of the binary system}, and is given by \citep{Zahn2008}:
\begin{equation}
\frac{dL_{\rm T}}{dt} = - \frac{\Omega -n}{t_{\rm diss}} M R^{2} \left( \frac{m}{M} \right)^{2} \left( \frac{R}{a} \right)^{6}, 
\label{tidal_torque}
\end{equation}
where $t_{\rm diss}$ is the characteristic timescale for the dissipation of the kinetic energy {of tidal flows} due to the {turbulent friction in convective regions and to thermal diffusion in radiative ones.
To simplify the problem, we choose here as a first step to consider only tidal dissipation in the convective envelope of the star \citep[e.g.][]{Mathis2015a}. This will be particularly relevant for the case of F-type stars studied in sec. \ref{ill_case} (see the corresponding discussion). We thus integrate the tidal torque $dL_{\rm T}/dt$ over the volume of the convection zone.}

If we differentiate Eq.~(\ref{tidal_torque}) with respect to the radius of the star, we obtain the average contribution of each shell that can be expressed as:
\begin{equation}
\frac{d}{dt} \left( \frac{dL_{\rm T}}{dr} \right) \simeq  -6 \left( \frac{\Omega -n}{t_{\rm diss}} \right)I \left(\frac{m}{M} \right)^{2} \left( \frac{r^{5}}{a^{6}} \right),
\label{der_tide_torque}
\end{equation}
where the moment of inertia $I = M R^{2}$  is assumed to be independent of the shell for our purposes because it enters into the expression for the tidal lag that we assume to be the same for all the shells inside the convection zone \citep[c.f.][]{Zahn2008}.  { To simplify our treatment, all tidal torques, including those of the dynamical tide (cf. Sect.~\ref{intro}), will be modeled using a constant modified tidal quality factor computed independently (see Eq. \ref{simplifiedtides}).}

From Eqs.~(\ref{der_tide_torque}) and~(\ref{ave_s1}), we derive $S_{1\rm T}$ by assuming that the torque is applied to the shell $[r_{c}, R]$. Thanks to the orthogonality of the radial eigenfunctions, the time-dependent coefficients are given by:
\begin{equation}
\beta_{0k}(t) = \frac{3}{4} \int_{r_{\rm b}}^{R} \rho \, {r'}^{4}(\tilde{S}_{1 \rm W} + \tilde{S}_{1 \rm T}) \zeta_{0k}({r'}) \, d{r'},  
\end{equation}
where the time dependence comes from the variation of the stellar angular velocity $\Omega$ that we can assume to be an average over the whole star without significantly increasing the error of our approximations. Moreover, we shall consider below the case when the timescales for the variation of $\Omega$, $n$, and $a$ are much longer than the diffusion timescales associated with the eigenvalues, i.e., $\lambda_{nk}^{-1}$, so that the stationary solutions as given by Eqs.~(\ref{long_timescale}) apply for $k>0$. For $k=0$, we obtain the variation of $\alpha_{00}$ by a simple integration (cf. Eq.~\ref{alphabeta0}), where $\beta_{00}$ is derived from the variation of the total angular momentum by means of Eq.~(\ref{tot_ang_mom_var}). 

From a physical point of view, the stationary solution is characterized by an angular velocity profile $\omega(r)$ whose gradient maintains a constant angular momentum flux across each spherical shell that balances the contributions coming from the tidal and wind torques acting inside the convection zone. Therefore, we expect that the variations of $\omega$ are localized not too far from the layers where the torques act, in particular where $\eta_{\rm eff}(r)$ is close to its minimum, because a larger gradient of $\omega$ is required to transport the angular momentum there. Far from those layers, we expect $\omega$ to be almost uniform and close to $\omega_{\rm mean}$ because the additional angular momentum coming from tides and winds is rapidly redistributed across the whole internal layers on a timescale shorter than the variation of the torques themselves. We shall verify those predictions in the following illustrative case {of F-type stars}. 

\section{An illustrative case: F-type stars hosting hot Jupiters}
\label{ill_case}

\subsection{Modeling stellar structure and evolution, internal stresses and applied torques}
\label{Model_Fstar}

As an illustrative application of the theory introduced above, we consider a binary system consisting of an F-type main-sequence star of mass $M =1.4$ M$_{\odot}$ and {a hot Jupiter of mass $m = 0.0035\, M$}.

The stellar interior model is computed by means of  the EZ-Web stellar evolution code\footnote{http://www.astro.wisc.edu/{$\sim$}townsend/static.php?ref=ez-web}. The chemical composition {is} assumed {to be} solar-like (i.e. $Z=0.02$, where $Z$ is the stellar metallicity). The model is evolved for 1.205 Gyr when the star reaches a radius $R=1.5576$~R$_{\odot}$ (cf. Fig.~\ref{radiusvstime}). 
\begin{figure}
\centerline{
 \includegraphics[width=13cm,height=9cm,angle=0]{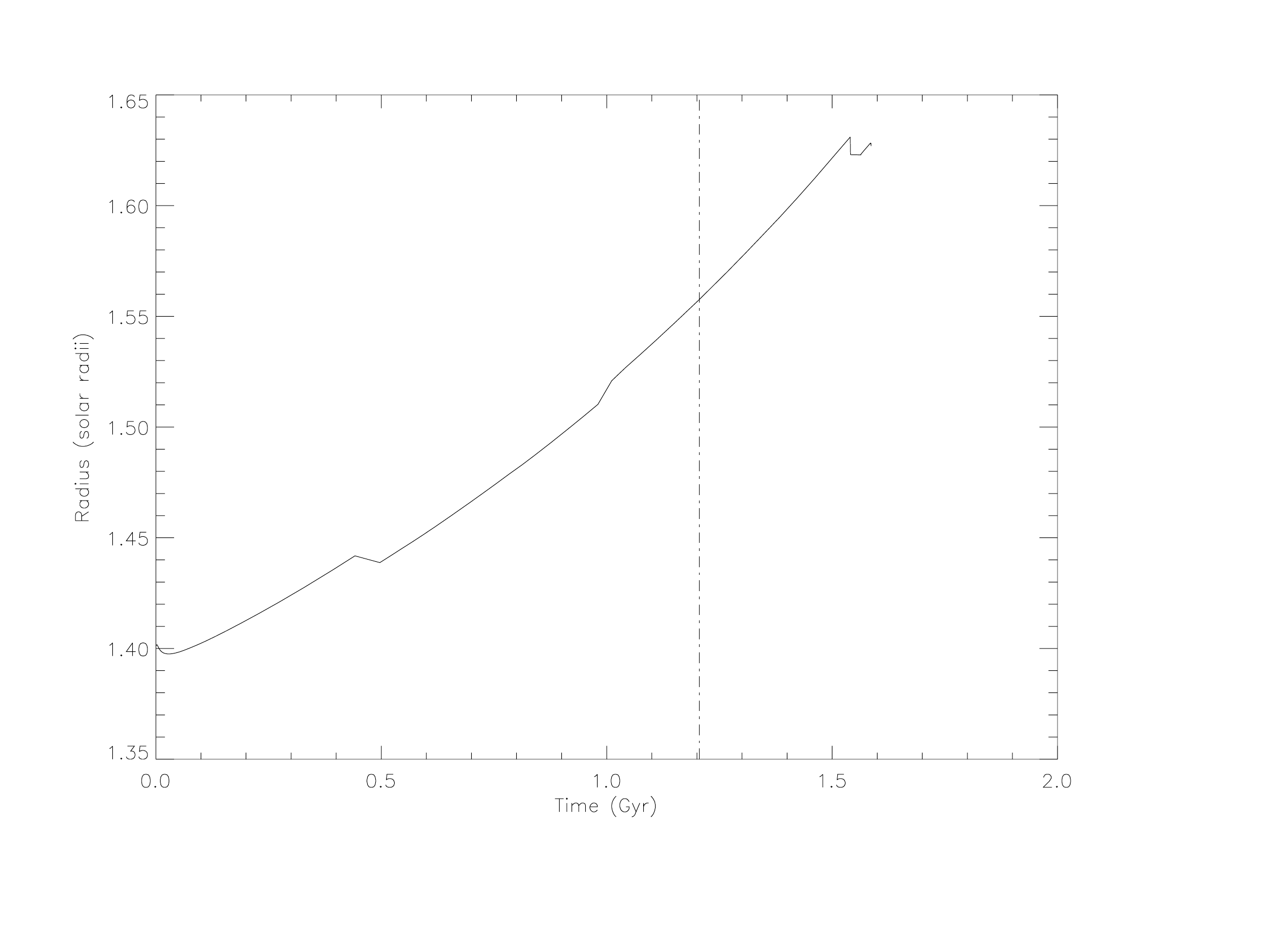}}
 \caption{The radius of our primary F-type star vs. the time since the Zero-Age Main-Sequence {(hereafter noted ZAMS)}. The dot-dashed line marks the age of 1.205~Gyr at which we consider its interior model. }
 \label{radiusvstime}
\end{figure}
The star has an inner convective core that extends up to $r_{\rm b}/R=0.0739$, while the outer convection zone extends down to $r_{c}/R = 0.9309$ with the radiative zone in between. The density {radial profile} in the stellar interior is plotted in Fig.~\ref{density_plot}.\\ 

In convective regions, the effective eddy-viscosity is evaluated using the usual mixing-length theory
\begin{equation}
\eta_{\rm eff}\left(r\right) = \frac{1}{3} \, \rho \, \alpha_{\rm mlt} \, H_{\rm p} \, v_{\rm conv},
\end{equation}
where $\alpha_{\rm mlt}=1.7$ is the ratio of the mixing length to the local pressure scale height $H_{\rm p}=\vert {\rm d}r/ {\rm d}\ln P\vert$ ($P$ being the pressure) and $v_{\rm conv}$ the {mean vertical} convective velocity given by the interior model {in this framework \citep[e.g.][]{BV1958,Zahn1989}}.
 
In the radiative zone, we impose the value of the kinematic {effective eddy-viscosity} $\nu_{\rm eff}$ in order to have an {effective} diffusion time {for the angular momentum} $\tau_{\rm eff} = 1$~Gyr across the zone itself, { if the spatial scale of the angular velocity variation is of the order of $r_{\rm c}$. Therefore, we fix $\nu_{\rm eff} = r_{c}^2/\tau_{\rm eff}$ and compute $\eta_{\rm eff} = \rho \nu_{\rm eff}$.  The actual timescale for the redistribution of the angular momentum,  given by the inverse of the lowest order eigenvalue, is indeed much shorter owing to the spatial variation of the associated eigenfunction on a scale smaller than $r_{\rm c}$ (see below).} {We recall that this effective (vertical) eddy-viscosity models here the (diffusive) action of the Reynolds stresses produced by the shear-induced turbulence and internal gravity waves \citep[see e.g.][]{TZ1997,MPZ2004,TC2005,Mathisetal2013,Rogers2015} that may possibly enforce a weak differential rotation as recently revealed by asteroseismology in some F-type stars observed with {\it Kepler} \citep{Kurtzetal2014,Saioetal2015,Benomaretal2015}.}  We here recall that in our simplified model, we neglect the potential anisotropy of the viscosity induced by the Coriolis acceleration and the Lorentz force, in view of the lack of a detailed non-linear theory of their effects \citep[cf.][]{Kitchatinovetal94} and of our ignorance of the internal magnetic fields of the stars. 

The effective eddy-viscosity is plotted in Fig.~\ref{eta_plot}, while an enlargement in an interval across the lower boundary of the outer convection zone is plotted in Fig~\ref{eta_plot_detail}. Note that $\eta_{\rm eff}$ reaches its minimum just above the lower boundary of the {convective envelope} because of the remarkable decrease of the convective velocity there as given by  the standard mixing length theory. This plot will be used to interpret the angular velocity variation vs. the radius in Sect.~\ref{results}. 
\begin{figure}[h!]
\centerline{
 \includegraphics[width=13cm,height=9cm,angle=0]{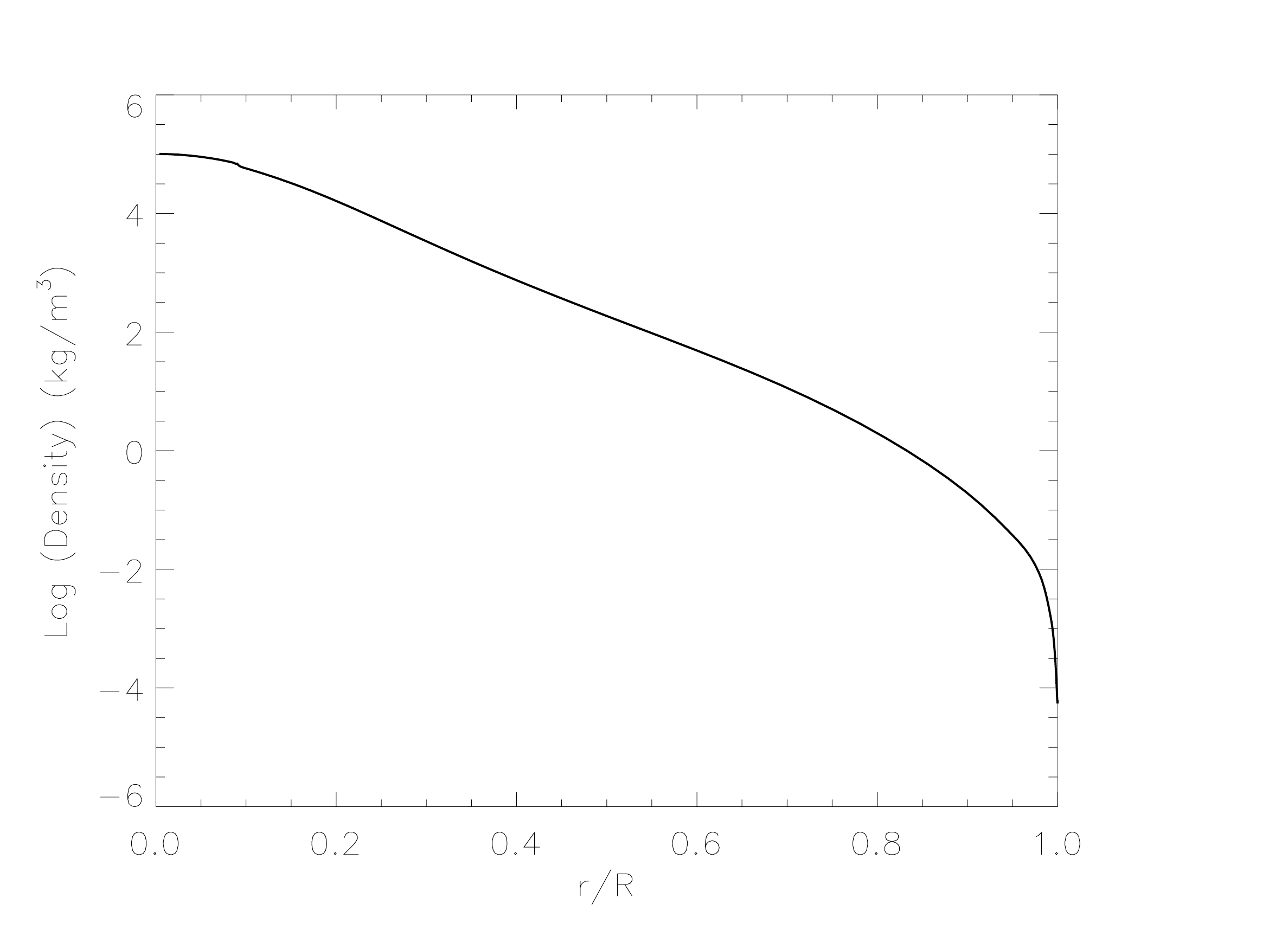}}
 \caption{The stellar density in the interior of our primary F-type star vs. the relative stellar radius. }
 \label{density_plot}
\end{figure}
\begin{figure}[h!]
\centerline{
 \includegraphics[width=13cm,height=9cm,angle=0]{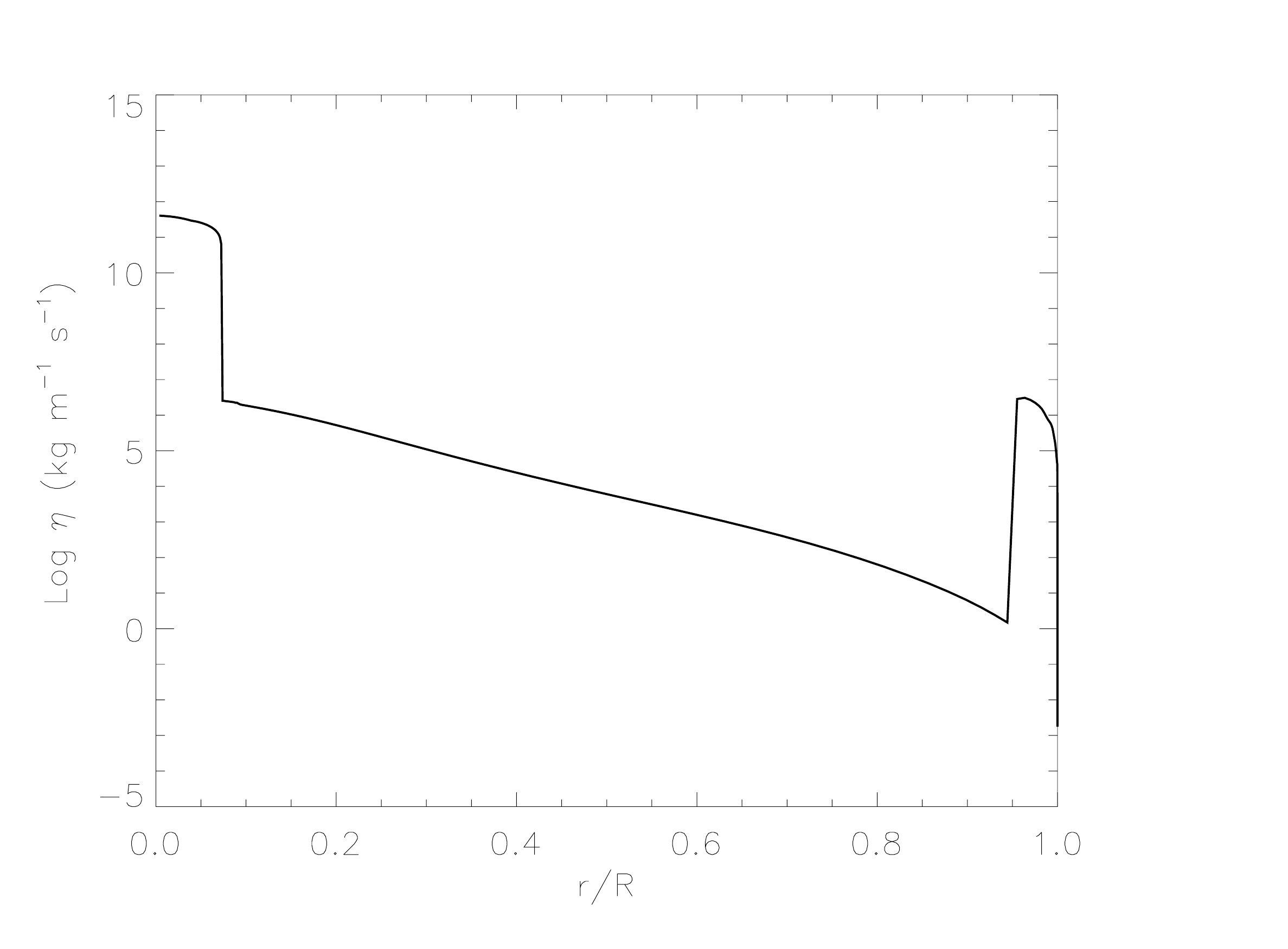}}
 \caption{The dynamical {effective eddy-}viscosity $\eta_{\rm eff}$ in the interior of our primary F-type star vs. the relative stellar radius. }
 \label{eta_plot}
\end{figure}
\begin{figure}[h!]
\centerline{
 \includegraphics[width=13cm,height=9cm,angle=0]{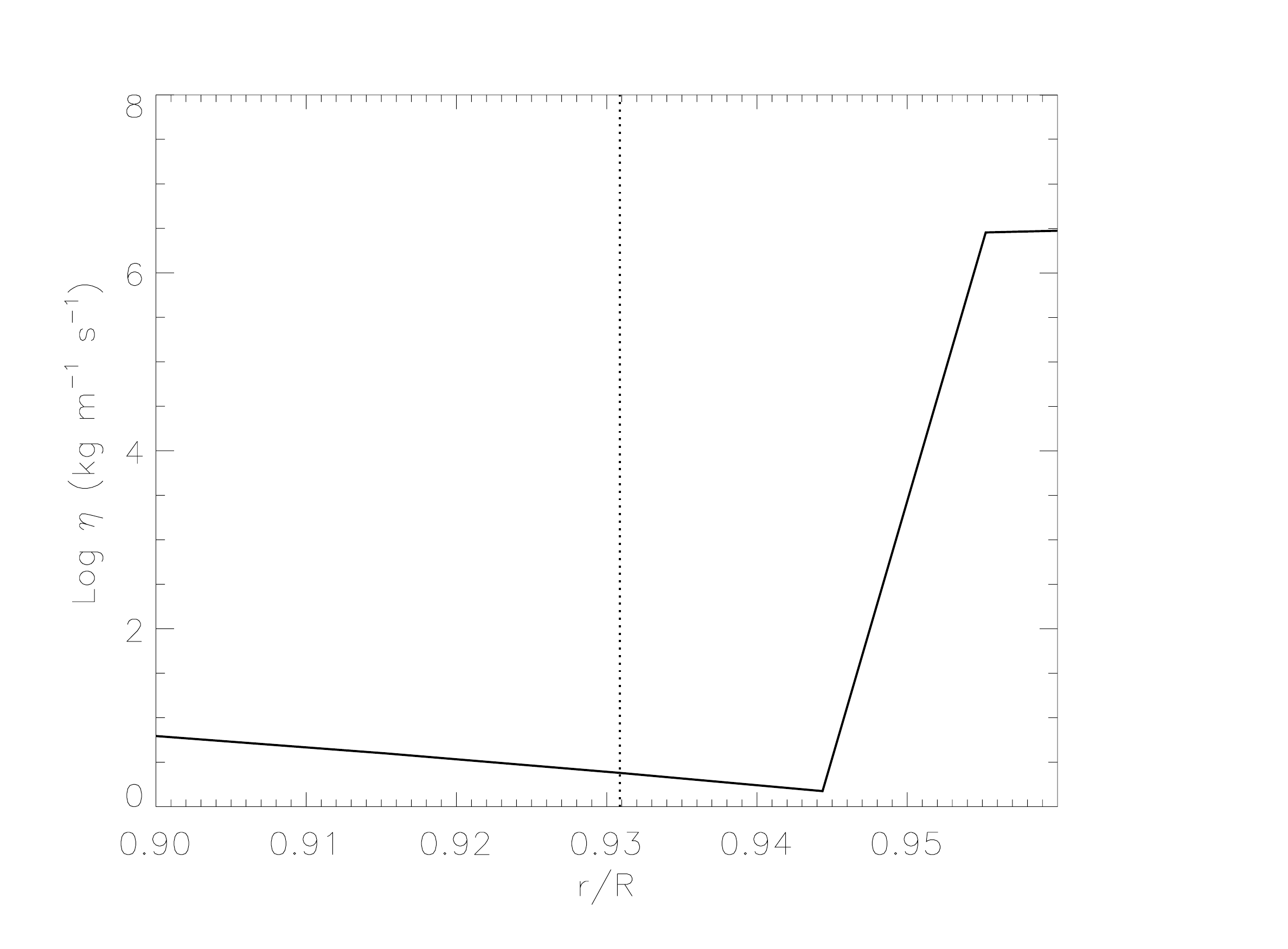}}
 \caption{The dynamical {effective eddy-}viscosity $\eta_{\rm eff}$ vs. the relative stellar radius close to the base of the outer convection zone of our F-type star. This plot is an enlargement of Fig.~\ref{eta_plot} in the radial interval of interest. The vertical dotted line marks the base of the convection zone where $v_{\rm conv}=0$.  }
 \label{eta_plot_detail}
\end{figure}

The eigenvalues and the eigenfunctions for the radial dependence of the angular velocity are computed by integrating Eq.~(\ref{radialpart}) from the stellar surface at $r=R$ inwards, down to the boundary of the inner convective core at $r=r_{\rm b}$, where we apply the boundary condition {given in Eq.} (\ref{bc1}) on the eigenfunction derivative. To find the eigenvalues, we use a bisection method, starting from a trial-and-error adjustment of the extrema of the interval to be searched and ending the search when the absolute value of the derivative at $r=r_{\rm b}$ becomes smaller than a prescribed (very small) value.  By counting the number of zeros of the eigenfunctions, we can be certain of identifying all the successive eigenvalues and eigenfunctions. The numerical integration is performed by means of the {\tt ODEINT} subroutine \citep{Pressetal1992}.\\ 

We compute {the dynamical evolution of our star-planet} system under the simplifying hypotheses that: i) {the stellar radius $R$ evolves as shown in Fig.~\ref{radiusvstime}, while the mass $M$ is constant}; ii) the angular momentum loss associated with the stellar wind is given by: 
\begin{equation}
\frac{dL_{\rm W}}{dt} = - K_{\rm wind} \Omega^{3};
\label{wind_am_loss_rate}
\end{equation}
and iii) the tidal torque applied onto the star is given by Eq.~(\ref{tidal_torque}) with an opposite torque applied to the orbit. 

{On the one hand, }the constant $K_{\rm wind} = 3.52 \times 10^{39}$~kg~m$^{2}$~s is held fixed at a value that is $\approx 10$ times smaller than in the case of a solar-like star because the angular momentum loss is lower in F-type than in G-type stars \citep[e.g.,][]{Lanzaetal2011}. 
{On the other hand,} the {tidal} dissipation time is held fixed along the evolution and is determined from the modified tidal quality factor $Q^{\prime}$ of the star{, which has been defined by \cite{OL2007}}, according to the parameterization {given by} \cite{Zahn2008}:
\begin{equation}
t_{\rm diss} = \frac{4}{9} \frac{[R(t_{0})]^{3}}{GM} | n(t_{0}) - \Omega(t_{0}) | Q^{\prime},
\label{simplifiedtides}
\end{equation}
where $n(t_{0})$ is the mean orbital motion and $\Omega(t_{0})$ the {global} spin angular velocity of the primary (noted $\omega_{\rm mean}$ in Sec. \ref{ill_case}) at the initial time $t_{0}$ {on the ZAMS}. {We recall that introducing $Q^{\prime}$ is a simple way to parametrize the tidal dissipation and that, as in the case of a damped oscillator, the lower $Q^{\prime}$ the stronger the tidal kinetic energy dissipation and the faster the evolution of the system towards its final configuration. Rigorously speaking, $Q^{\prime}$ {may vary over several orders of magnitude as a function of the age, rotation, stratification, viscous and thermal diffusivities, and tidal frequency for a given stellar mass \citep[e.g.][]{OL2007,ADMLP2015,Mathis2015a}. 

Studying the case of F-type stars with $M>1.1M_{\odot}$ (i.e. with a convective core) allows us to do several simplifications. First, we can neglect tidal dissipation in the convective core. Indeed, tidal inertial waves excited there are regular modes leading only to a weak dissipation in opposition to singular ones excited in the external convective envelope that lead to sheared waves attractors where strong dissipation may occur \citep[e.g.][]{Wu2005,BO2009,Ogilvie2013}. { Next, since our goal is to compute only an illustrative example for our method, we ignore as a first step the dissipation of linear tidal internal gravity waves excited in the radiative zone. Note however, that this dissipation can be comparable or larger than the tidal turbulent friction in the thin external convective envelope as pointed out by \cite{ValsecchiRasio14}. They estimate that tidal internal gravity waves may dissipate at least 1-2 orders of magnitude more power than the equilibrium tide in the case of an F-star with a mass of $\sim 1.5 M_{\odot}$ when the star's spin frequency is sufficiently away from the orbital frequency as in most of non-synchronized star-planet systems.} Moreover, as pointed out by \cite{Guillotetal2014}, the presence of this core may inhibit the possible nonlinear breaking of { high-amplitude} tidal gravity waves excited and propagating in the radiative zone as expected in lower-mass stars because of their central focusing \citep[][]{BO2010}. We thus focus only on the dissipation of tidal inertial waves propagating in the external convective envelope. To simplify the problem for this first work, we choose to consider an equivalent modified tidal quality factor corresponding to the frequency-averaged dissipation  \citep[e.g.][]{Ogilvie2013,Mathis2015a,Mathis2015b}. This latter depends only on stellar structure and rotation  and allows us to filter out the complex frequency dependence of the dissipation of tidal (gravito-)inertial waves \citep[e.g.][]{OL2007} in contrast to the smooth frequency variation of the dissipated power in the usually adopted weak-friction approximation. Moreover, the extraction of angular momentum by stellar winds in F-type stars is relatively weak as discussed above, while the structure of the star is almost fixed during its main-sequence (hereafter noted MS) evolution. Therefore, we can} assume that $Q^{\prime}$ and $t_{\rm diss}$ {are} constant throughout the tidal evolution {studied here along the MS} and equal to {their} initial values {on the ZAMS} as obtained for a rotation period of the star $P_{\rm rot} (t_{0}) = 6$~days and a  semimajor axis of the orbit $a(t_{0}) = 0.05$~AU.  

We  integrate the evolution equations for the stellar and orbital angular momenta. Specifically, we solve the equation for the semimajor axis of the orbit
\begin{equation}
\frac{da}{dt} = -2 \frac{\sqrt{a (M+m)}}{\sqrt{G} M m} \frac{dL_{\rm T}}{dt}, 
\label{eq_evol1}
\end{equation}
where $dL_{\rm T}/dt$ is given by Eq.~(\ref{tidal_torque}) with $\Omega = \omega_{\rm mean}$, and that for the mean stellar angular velocity (cf. Eq.~\ref{wind_am_loss_rate}) 
\begin{equation}
\frac{d\omega_{\rm mean}}{dt} =\frac{1}{I_{*}}\left( \frac{dL_{\rm T}}{dt} - K_{\rm wind} \omega_{\rm mean}^{3} \right),
\label{eq_evol2}
\end{equation}
where $I_{*}$ is the moment of inertia of the star assumed to be virtually equal to that of the shell $I_{\rm c}$ considered in our model.  It is computed taking into account the variation of the stellar radius, i.e.,  $I_{*} = M [g_{\rm r} R(t)]^2$, where $g_{\rm r} = 0.35$ is the relative gyration radius of  the star that is assumed constant.  

We integrate Eqs.~(\ref{eq_evol1}) and~(\ref{eq_evol2}) starting from the above initial configuration with the star on the ZAMS ($t=t_{0}$) up to the time $t=1.205$~Gyr. To perform the numerical integration of Eqs.~(\ref{eq_evol1}) and~(\ref{eq_evol2}), we use {\tt ODEINT} again. Since this is a {first} illustrative calculation, we {compute the evolution of the global stellar angular momentum as if the star is rotating as a solid body to simplify its calculation and those of $a$. In this framework, we introduce the associated mean angular velocity of the star $\omega_{\rm mean}$}. Of course, this is not consistent with the results on the internal rotation, where a radial gradient of the angular velocity is found to be present, but we postpone a fully consistent evolutionary calculation to a successive study because the purpose of the present work is only to {make a proof of concept of our} mathematical {dynamical model. It allows for the first time to compute the simultaneous evolution of the orbital angular momentum of the planet and of the spin and internal radial differential rotation of its host star.}

\subsection{Results}
\label{results}
{As discussed in previous sections, we focus here on the} latitudinally averaged model {introduced} in Sect.~\ref{lat_ave_model}. 
The first 30 eigenvalues as obtained from the numerical solution of the Sturm-Liouville problem for $n=0$ are listed in Table~\ref{table1} in units of the inverse of the diffusion time we assumed for the radiative zone $\tau_{\rm diff}^{-1}$. We recall that the diffusion of  the angular momentum associated with the $k$-th eigenfunction occurs with a characteristic timescale $\lambda_{nk}^{-1}$ that decreases rapidly as $\approx k^{-2}$ for $n=0$. In our case, the slowest decaying mode with $k=1$ has a characteristic diffusion time $\lambda_{01}^{-1} \approx 53$~Myr. { This is much shorter than the timescales of the variation of the background stellar rotation $\Omega_{0}$ under the effects of the wind braking and the tides, that was a requisite for the validity of our approximation (cf.~sec~\ref{assumptions}). Note that $\lambda_{01}^{-1}$ is comparable with the characteristic timescale of $10^{7}-10^{8}$~yr for the angular momentum transport by internal gravity waves and magnetic fields estimated for solar-like stars \citep{Zahnetal97,Denissenkovetal08}, that supports our assumption for $\tau_{\rm eff}$.  }

The first five normalized eigenfunctions obtained for $n=0$ are plotted in Fig.~\ref{eigen_plot}. Their amplitude variations can be understood on the basis of Eq.~(\ref{zeta_asymp}) because $\sqrt{\rho/\eta_{\rm eff}}$ is much larger in the radiative zone than in the outer {convective envelope}, thus producing a remarkable decrease of their amplitude there.  The decrease of the density with radius also plays a role, making their amplitude larger towards the outer layers of the radiative zone. 
\begin{table}
\caption{Eigenvalues of the Sturm-Liouville problem as defined by Eqs.~(\ref{radialpart}) and (\ref{bc1}) for $n=0$. They are obtained by numerical integration as detailed in sec.~\ref{ill_case}.  The zeros of the $k$-th eigenfunction are $k-1$. The diffusion time is $\tau_{\rm diff} = 1$~Gyr. }
\begin{center}
~\\
\begin{tabular}{rrrrrr}
\hline
& & & & & \\
$k$ & $\lambda_{0k}$ & $k$ & $\lambda_{0k}$ & $k$ & $\lambda_{0k}$ \\
 & ($\tau_{\rm diff}^{-1}$) & & ($\tau_{\rm diff}^{-1}$) & & ($\tau_{\rm diff}^{-1}$) \\
 & & & & & \\
 \hline
  & & & & & \\
 1 & 18.828 & 11 & 1314.45 & 21 & 4814.06 \\
 2 & 52.302 & 12 & 1562.89 &   22 & 5285.94 \\ 
 3 & 110.11 & 13 & 1832.81 & 23 & 5784.37 \\
 4 & 188.44 & 14 & 2126.17 & 24 & 6306.23 \\
 5 & 286.68 & 15 & 2442.00 &  25 & 6845.31 \\
 6 & 404.84 & 16 & 2779.37 & 26 & 7415.63 \\
 7 & 544.34 & 17 & 3142.66 & 27 & 8008.59 \\
 8 & 704.37 & 18 & 3525.00 & 28 & 8591.41 \\
 9 & 885.94 & 19 & 3932.42 & 29 & 9259.38 \\
 10 &    1089.06 & 20 & 4362.50 & 30 & 10578.1\\
  & & & & & \\ 
\hline
\label{table1}
\end{tabular}
\end{center}
\end{table}
\begin{figure}
\centerline{
 \includegraphics[width=13cm,height=9cm,angle=0]{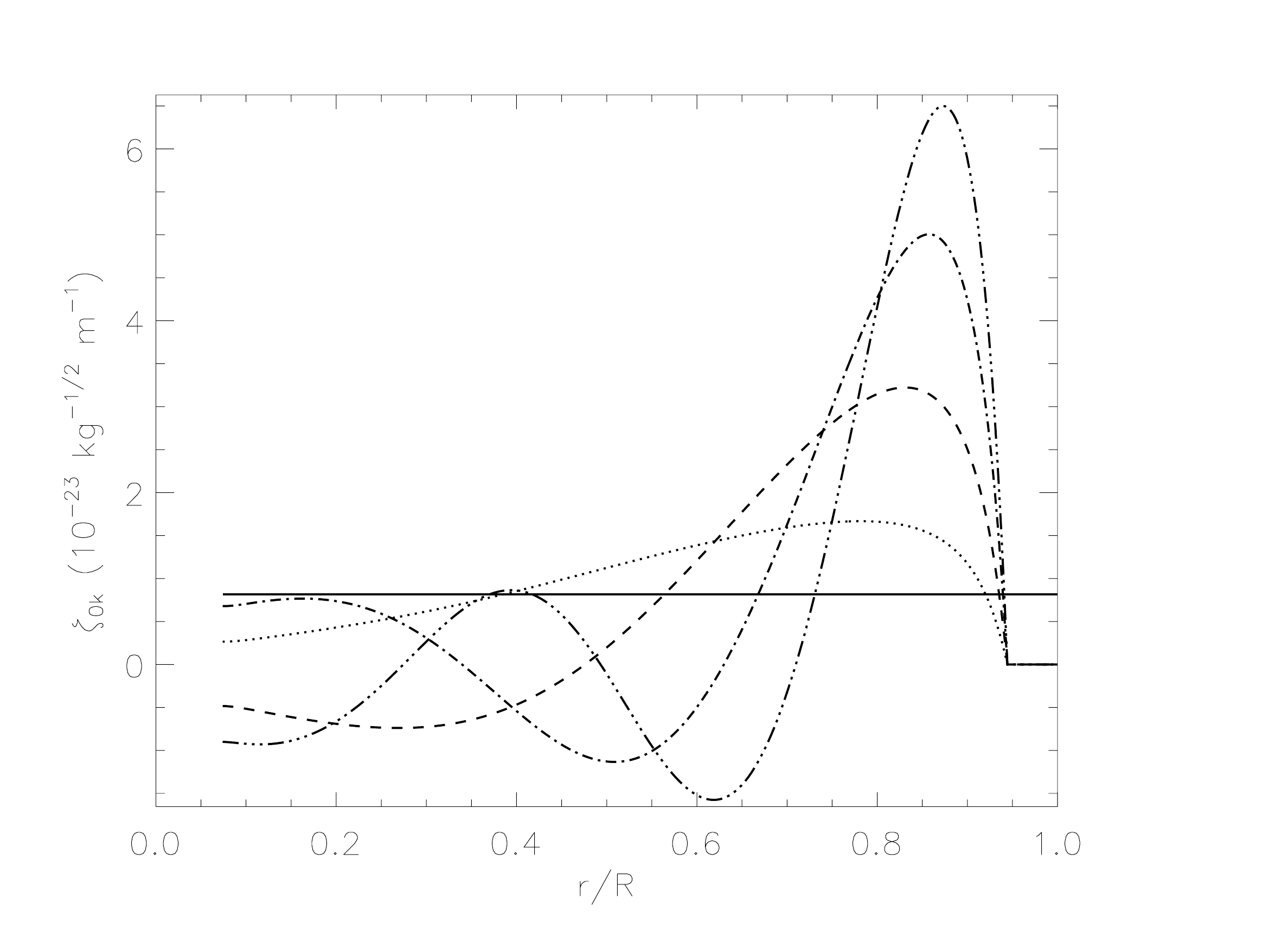}}
 \caption{The first five normalized eigenfunctions $\zeta_{nk}$ for $n=0$ vs. the relative radius. Different linestyles mark  different values of $k$. The eigenfunction with $k=0$ is associated with the zero eigenvalue and is constant (solid line). The other eigenfunctions are those with $k=1$ (dotted line); $k=2$ (dashed line); $k=3$ (dash-dotted line); and $k=4$ (dashed-three-dotted line). Note that the eigenfunction of order $k$ has $k-1$ radial zeros and that the starting value for the integration at $r=R$ is the same for all the unnormalized eigenfunctions.  As a consequence of the large increase of the {effective eddy-}viscosity $\eta_{\rm eff}$  in the outer convection zone, all the eigenfunctions are very small there. }
 \label{eigen_plot}
\end{figure}

The evolution of the spin angular velocity $\omega_{\rm mean}$ and orbital semimajor axis $a$ of our binary system is plotted in Fig.~\ref{tidal_evolution} for $Q^{\prime} = 7 \times 10^{6}, 10^{7}$, and $2 \times 10^{7}$.  The star is initially rotating slower than the mean orbital motion, therefore tides transfer angular momentum from the orbit to the stellar spin producing a decrease of the semimajor axis. The loss of angular momentum due to the stellar wind is not able to counteract the action of tides {for $Q^{\prime} < 2 \times 10^{7}$,}  thus the stellar angular velocity is increasing vs. the time, {except in the third case when the two torques are opposite and almost equal, so it stays almost constant}. The orbital angular momentum is smaller than three times the stellar spin angular momentum, so the system is tidally unstable and the planet fate is to fall into the star without reaching synchronization between the stellar spin and the orbital motion.\\ 
\begin{figure}
\centerline{
 \includegraphics[width=13cm,height=9cm,angle=0]{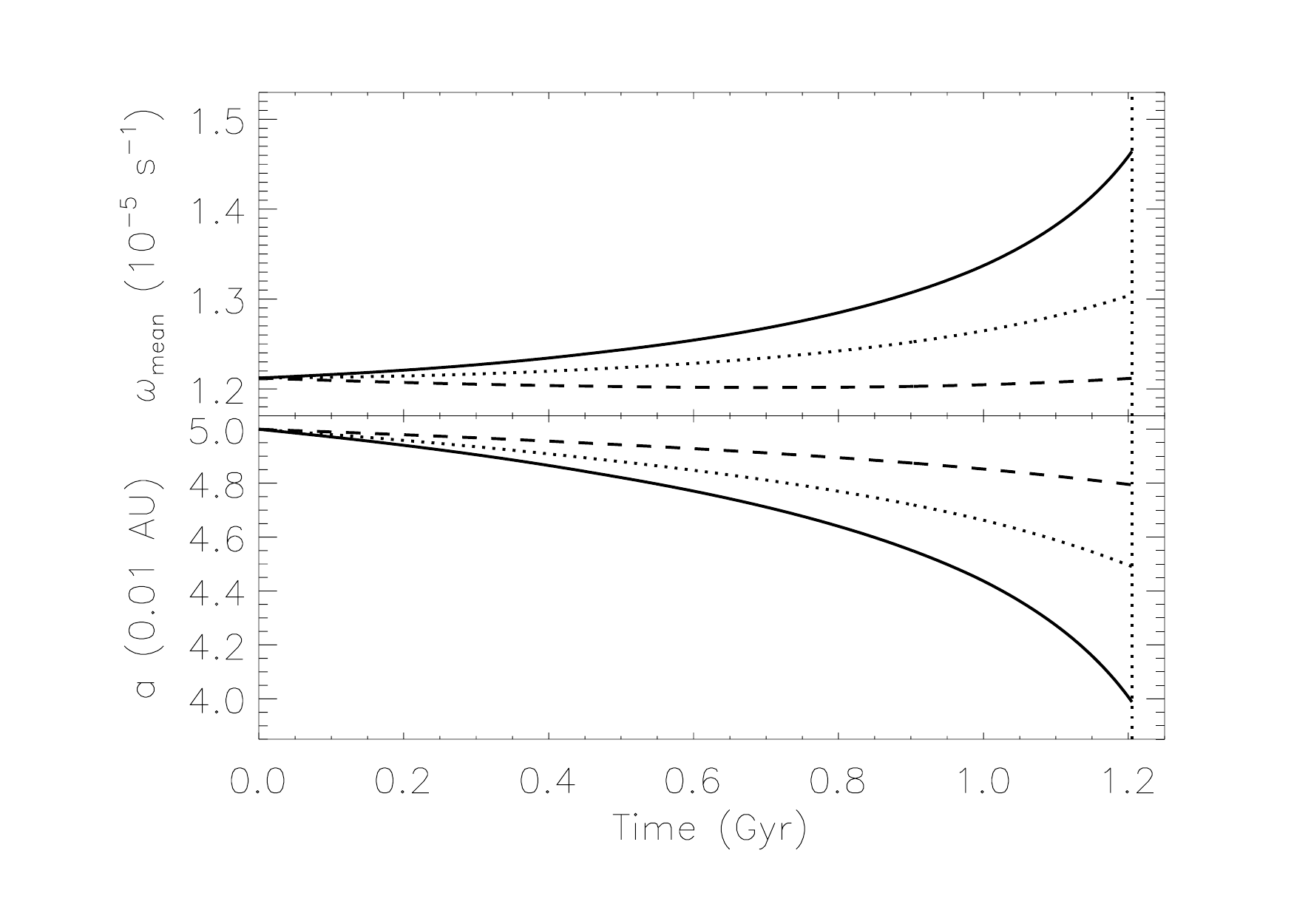}}
 \caption{Upper panel: the stellar spin angular velocity $\omega_{\rm mean}$ vs. the time for three different values of the modified stellar tidal quality factor $Q^{\prime}$: $7 \times 10^{6}$ (solid line); $10^{7}$ (dotted line); $2 \times 10^{7}$ (dashed line). Lower panel: the orbit semimajor axis $a$ vs. the time. Different linestyles indicate different values of $Q^{\prime}$ as in the upper panel. The vertical dotted line marks the epoch corresponding to the internal angular velocity plotted in Fig.~\ref{solutions} when we stop computing the evolution.}
 \label{tidal_evolution}
\end{figure}

The internal torques produced by the stellar wind and tides for $Q^{\prime}=7 \times 10^{6}$ are plotted vs. the radius in the convection zone in Fig.~\ref{internal_torques}. The wind torque is negative, while the tidal torque is positive because the mean orbital motion is faster than stellar rotation. In this case, the tides overcome the wind angular momentum loss  leading to an acceleration of the internal rotation in a certain layer. 
\begin{figure}
\centerline{
 \includegraphics[width=13cm,height=9cm,angle=0]{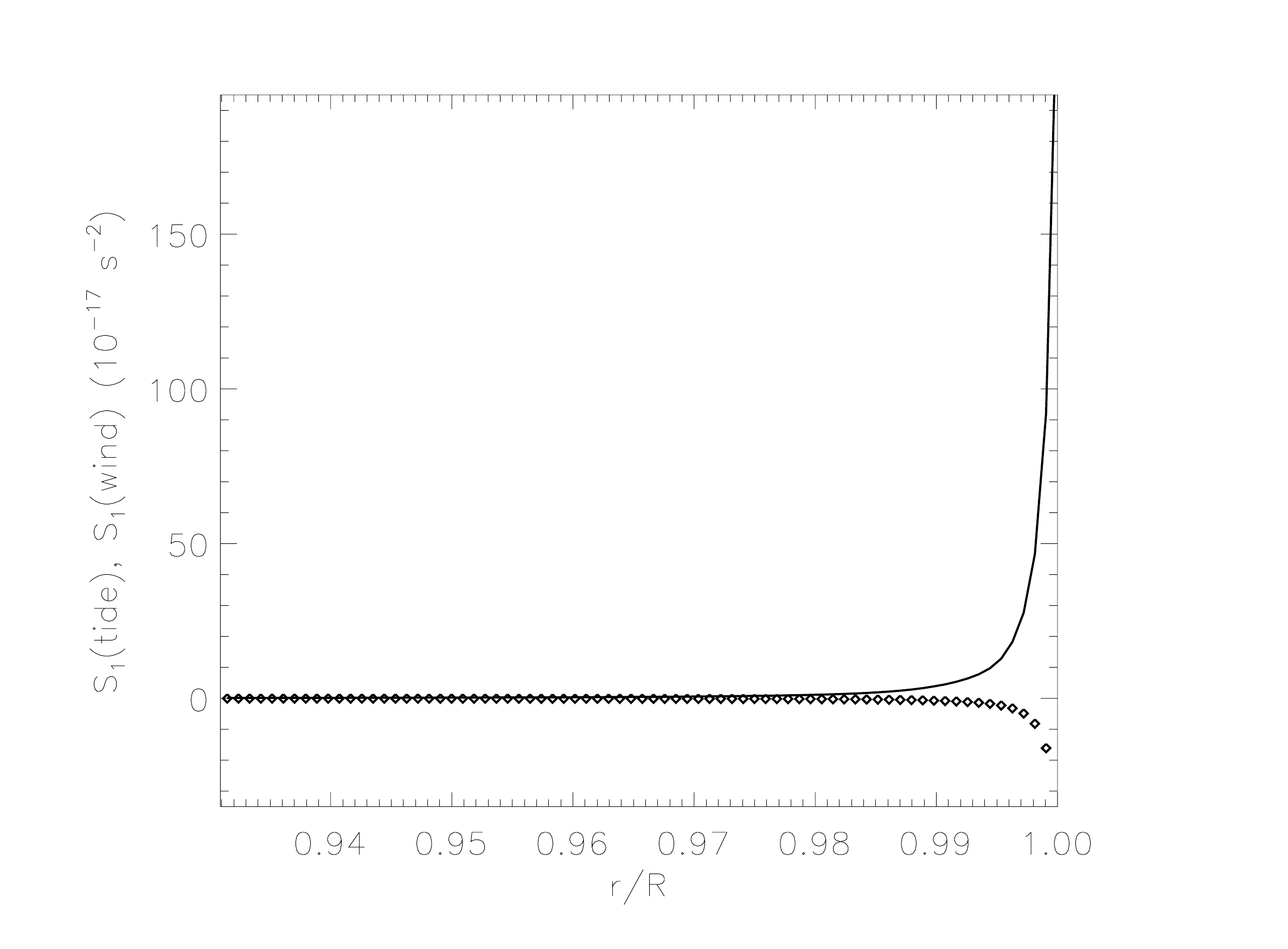}}
 \caption{The angular velocity source term $\tilde{S}_{1}$ produced by the stellar wind torque (diamonds) and by the tidal torque (solid line) vs. the radius in the outer convection zone, respectively. These are latitudinally averaged source terms as introduced in Sect.~\ref{lat_ave_model} {computed for $Q^{\prime} = 7 \times 10^{6}$ and $t-t_{0}= 1.205$~Gyr}. }
 \label{internal_torques}
\end{figure}

Indeed, the latitudinally-averaged stellar angular velocity at $t-t_{0}=1.205$~Gyr is plotted in Fig.~\ref{solutions} vs. the relative stellar radius  for the three different values of the modified stellar tidal quality factor $Q^{\prime}$. We model the rotation inside the radiative zone and the outer convection zone (cf. Sect.~\ref{lat_ave_model}) and discard the inner convective core because we find that the variation of the angular velocity is localized in the upper layers of the radiative zone, while $\omega$ becomes virtually indistinguishable from its mean value $\omega_{\rm mean}$ in its lower part. Therefore, we expect that $\omega$ remains very close to $\omega_{\rm mean}$ also in the inner convective core because its moment of inertia is much smaller than that of the layers above (see below). {Moreover}, note that asteroseismic techniques {would not be} sensitive to rotation gradients too close to the centre {of solar-type stars on the MS because of the lack of detection of individual gravity (g-) modes \citep[see e.g. the detailed discussion in ][and references therein]{Benomaretal2015,Mathuretal2008}}. The displayed solutions are computed by considering the first 40 eigenfunctions as obtained by numerically solving the eigenvalue problem for $n=0$. Although the eigenvalues computed by means of the asymptotic formula (\ref{eigen_asymp}) differ by only 1.8 percent from those obtained by the numerical solution for $k>20$, the corresponding eigenfunctions do not provide a solution of quality comparable with that obtained from the eigenfunctions computed by  numerical integration. This is likely a consequence of the strong gradients in $\eta_{\rm eff}(r)$ present in our model that make the asymptotic expressions  vary too much in comparison to their true expressions. We estimate that only for $k> 50-60$ the asymptotic eigenfunctions provide results comparable to those obtained with the eigenfunctions computed by integrating {numerically} the radial equation. Our solutions are computed by Eq.~(\ref{slow_solution}), i.e., by assuming that the variation of the time-dependent coefficients $\beta_{0k}$ is much slower than the characteristic timescales of angular momentum diffusion $\lambda_{0k}^{-1}$, the longest of which is $\approx 53$~Myr (cf. Table~\ref{table1}). Moreover, the solutions are computed at a time $t-t_{0}$ much longer than all the $\lambda_{0k}^{-1}$, therefore, they get very close to the stationary state as given by Eq.~(\ref{long_timescale}). 

For $Q^{\prime} = 7 \times 10^{6}$ (Fig.~\ref{solutions}, solid line), the angular velocity shows a remarkable increase close to the base of the convection zone in the interval where $\eta_{\rm eff}$ reaches its minimum (cf. Fig.~\ref{eta_plot_detail}).  This is due to the tidal torque that is applied in the convection zone and provides more angular momentum than can be removed by the wind. {Moreover, in the interval where $\eta_{\rm eff}$ reaches its minimum, the transport is not efficient enough to erase the radial gradient of angular velocity induced by tides. In other words, a large local gradient of the angular velocity develops there to allow a transfer of the angular momentum towards the layers below, in spite of 
the locally small value of $\eta_{\rm eff}$.} This explains the steep increase of $\omega$ with the decrease of the radius. After reaching a relative maximum, the angular velocity decreases in a slower way towards the bulk of the radiative zone because  $\eta_{\rm eff}$ increases making the transport of angular momentum more efficient there, thus {leading to} a smaller gradient of $\omega$. The variation in $\omega$  is confined down to $\approx 0.4R$ with the angular velocity below that limit  virtually equal to the mean angular velocity $\omega_{\rm mean}$. This happens because $t-t_{0} \gg \lambda_{01}^{-1}$, so there has been enough time to redistribute the angular momentum in those layers making the {latitudinally-averaged} rotation uniform in that part of the star and practically indistinguishable from the {mean} value. The angular velocity approaches again the mean value $\omega_{\rm mean}$ in the bulk of the outer convection zone because the {effective eddy-}viscosity is large enough to effectively transport the angular momentum, even if the gradient of $\omega$ is vanishingly small. Therefore, we do not see any appreciable radial gradient of the angular velocity there. 

For $Q^{\prime} = 10^{7}$ (Fig.~\ref{solutions}, dotted line), {the angular velocity has a similar behaviour, although with a smaller amplitude because the unbalance between the tidal and the wind torques is smaller. For $Q^{\prime} = 2 \times 10^{7}$ (dashed line), our solution displays a layer with a slightly negative variation of the angular velocity across the base of the outer convection zone. This is a consequence of the wind torque that counterbalances the tidal torque, thus reversing the sign of the variation, while  the mean stellar angular velocity stays almost constant (cf. Fig.~\ref{tidal_evolution}, upper panel)}.  

Our solutions display  wiggles that are likely associated with the finite accuracy of the numerically computed eigenvalues and eigenfunctions as well as the truncation of our series at ${ k=40}$. By considering a truncation at $k=20$, we estimate that our solution is accurate at the level of { $1.5-3$ percent  of $\omega_{\rm mean}$  away from the sharp peak close to the base of the convection zone. This is comparable with the amplitude of the wiggles we observe. To further investigate the effects of the series truncation, we compute numerically the stationary solution of Eq.~(\ref{lataveeq}) in the case of $Q^{\prime} = 7\times 10^{6}$ by  applying a shooting method  to satisfy the boundary conditions. In Fig.~\ref{solutions_cp}, we compare this exact solution with those obtained by truncating the eigenfunction series at $k=25$ and $k=40$, respectively. The series with $k=40$ provides a good approximation, except close to the top of the sharp peak near the base of the convection zone because its radial resolution is of the order of $R/k \approx 0.025 R$ that is not adequate to reproduce that localized variation.  The differences are smaller in the cases with $Q^{\prime} = 10^{7}$ and $Q^{\prime} = 2 \times 10^{7}$ because of the less sharp variations. }
\begin{figure}
\centerline{
 \includegraphics[width=9cm,height=13cm,angle=90]{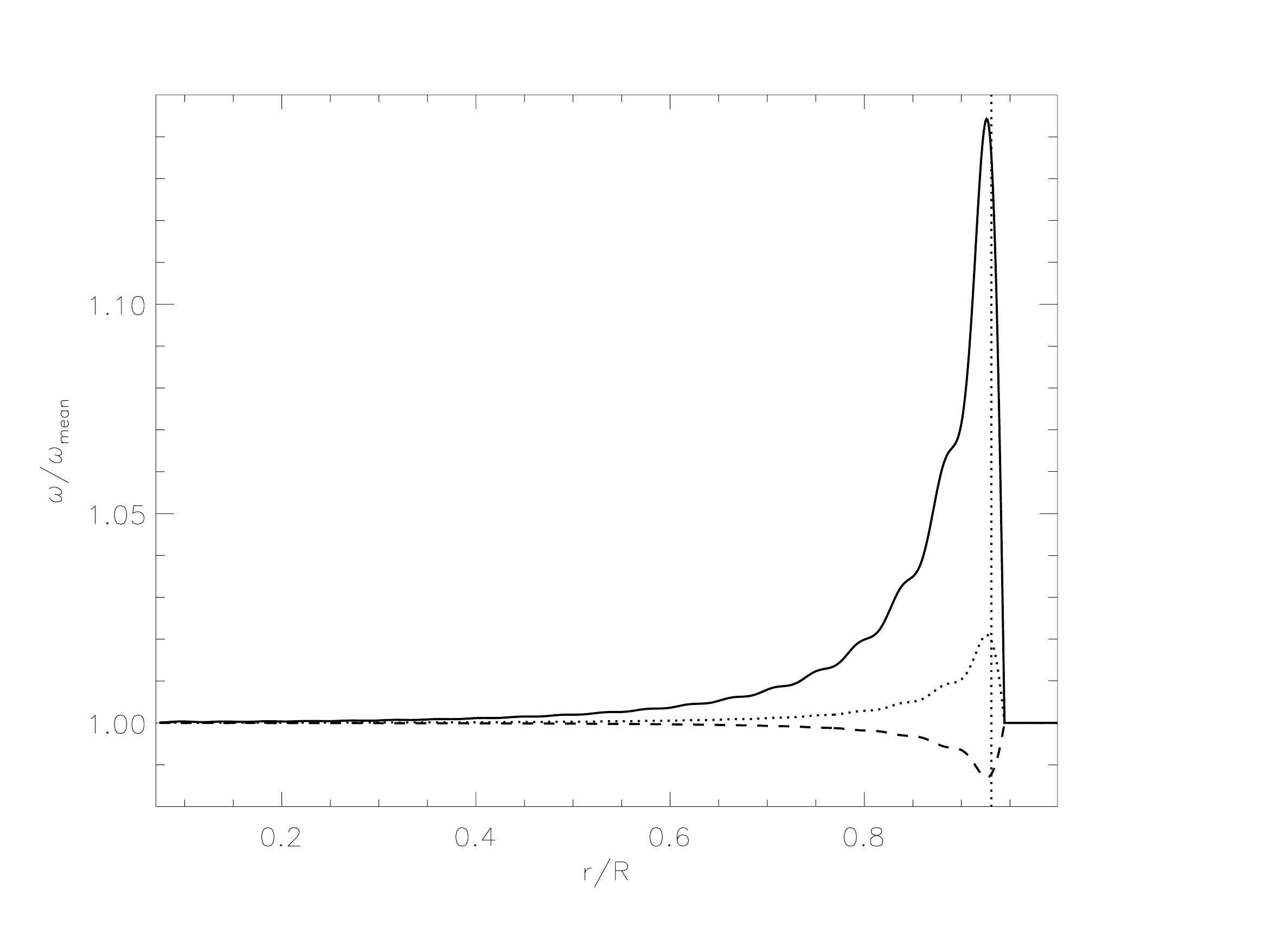}}
 \caption{Variation of the stellar angular velocity $\omega$ in units of $\omega_{\rm mean}$ vs. the relative stellar radius for three different values of the modified stellar tidal quality factor $Q^{\prime} =  7 \times 10^{6}$, solid line; $10^{7}$, dotted line; $2 \times 10^{7}$, dashed line. The mean angular velocity $\omega_{\rm mean}$ is plotted in the upper panel of Fig.~\ref{tidal_evolution} vs. the time. The present plots correspond to an age of $t-t_{0} = 1.205$~Gyr. The vertical dotted line marks the lower boundary of the outer stellar convective zone at $r=r_{\rm c}$. }
 \label{solutions}
\end{figure}
\begin{figure}
\centerline{
 \includegraphics[width=9cm,height=13cm,angle=90]{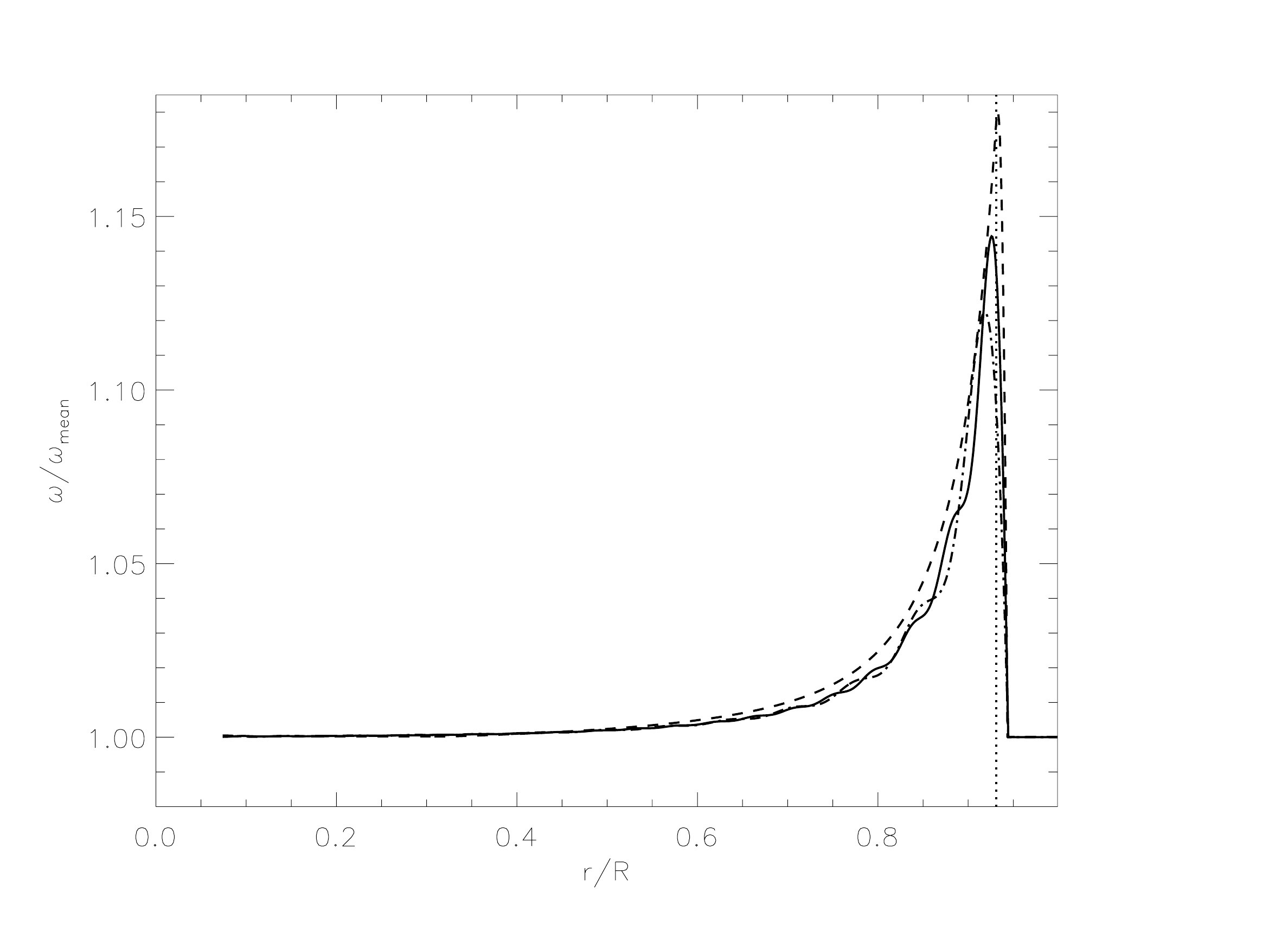}}
 \caption{  Variation of the stellar angular velocity $\omega$ in units of $\omega_{\rm mean}$ vs. the relative stellar radius for  $Q^{\prime}$ = $7 \times 10^{6}$ computed by a solving Eq.~(\ref{lataveeq}) with a numerical shooting method (dashed line) and with our eigenvalue series truncated at $k=25$ (dash-dotted line) and $k=40$ (solid line) to illustrate the impact of the eigenseries truncation on the computed solution. }
 \label{solutions_cp}
\end{figure}

\section{Discussion and conclusions}
\label{conclusions}
We introduced a {first dynamical} model to compute {simultaneously the global rotational evolution and the latitudinally-averaged differential rotation for} a late-type star under the {combined} effects of the torques associated with the stellar wind and the tides produced by a close-by companion {and the orbital evolution of this latter}. Our model is based on several simplifying assumptions among which the most {important} are: no angular momentum transport by internal magnetic fields or/{and} meridional flows; an internal {effective eddy-}viscosity depending only on the radius; and a negligible deformation of the stellar density stratification by rotation and tides. We present the mathematical basis of our approach that has the advantage of solving the equation for the internal angular momentum transport in an analytic way. {This model opens the path to the integrated study of the dynamical evolution of stars and of their surrounding (planetary) systems in a context where high-resolution photometric space missions such as CoRoT \citep{Baglinetal2006}, {\it Kepler} \citep{Boruckietal2010} and TESS \citep{Rickeretal2014} and PLATO \citep{Raueretal2014} in a near future are aimed to simultaneously characterize planets using the transit method and their host stars thanks to asteroseismology \citep[e.g.][]{Gizonetal2013,Daviesetal2015,Ceillieretal2016}. Thanks to its simplicity, our model will} be used {in forthcoming works to} test more sophisticated hydrodynamical numerical models {including realistic treatments of tidal flows and of their dissipation \citep[e.g.][]{Zahn1975,TK1998,OL2007,Ivanovetal2013,BR2013,Favieretal2014,Gueneletal2016}}.\\

As an illustrative case, we apply our model to  an F-type main-sequence star with a massive close-in planet {orbiting in its equatorial plane. The model is used} to compute the wind and tidal torques inside the {convective envelope} of the star. From the expressions of the torques averaged over the latitudes, we compute the variation of the internal angular velocity with the radius at a given epoch along the evolution of the system.  For simplicity sake, we assume a {slow} time variation of the external torques so that the stellar rotation profile can adjust itself to a stationary state.The deviation of the angular velocity from a uniform rotation is localized in the upper layers of the radiative zone, close to the base of the outer convection zone of the star, reaching  maximum values between $\sim 2$ and $\sim 15$  percent {of the mean rotation rate}, according to the assumed strengths of the tidal and wind torques. 

{ The modified tidal quality factor $Q^{\prime}$ we consider ranges between $7 \times 10^{6}$ and $2 \times 10^{7}$ to illustrate the transition from a tidal torque  stronger than the wind torque to a weaker one for the assumed mass of the hot Jupiter companion. Nevertheless, the recent calculations by \cite{Mathis2015b} suggest that $Q^{\prime}$ is higher by $\approx 3$ orders of magnitude owing to the shallow convective envelopes of F-type stars. In order to reproduce the same transition regime with a higher $Q^{\prime}$, it suffices to increase the mass $m$ of the secondary component because the tidal torque is proportional to $(m/M)^2$. For instance, considering a brown dwarf of mass $0.035\,M_{\odot}$, we increase the torque by a factor of $10^{2}$ approaching the stationary states considered above.} { It is important to remind here that our results constitute a first illustrative application of our method, which is only a first step because of the rough modeling of tidal dissipation mechanisms using the modified tidal quality factor. This should be improved in a near future by taking into account all dissipative mechanisms both in the radiative and convective regions and their frequency-dependence \citep[e.g.][]{OL2007,BO2009,ValsecchiRasio14}}

Although the gradients in the angular velocity we have found are quite localized and cannot be resolved with current asteroseismic techniques, it is hoped that they can become detectable with future methods, thus allowing us to directly probe the effects of the stellar wind and tides on stellar rotation in star-planet systems. The angular velocity gradients predicted by our model could also play a role in the internal mixing close to the base of the convection zone that affects the abundance of light elements such as Lithium, Boron, or Berillium \citep[e.g.,][]{Bouvier2008}. {Moreover,} differential rotation is key to the stellar hydromagnetic dynamo because it produces strong toroidal fields from weak poloidal ones in active late-type stars \citep[e.g.][and references therein]{Charbonneau2010,Brun2014}.  In this context, the  case of WASP-18, an F-type star with an anomalously low level of magnetic activity \citep{Pillitterietal2014}, could be explained as a consequence of its massive hot Jupiter that may modify its radial differential rotation close to the base of its outer convection zone where the large-scale dynamo is supposed to operate \citep{Augustsonetal2013}. If the differential rotation is decreased below a threshold value, turbulent {Ohmic} diffusion overcomes field amplification and the large-scale dynamo immediately shuts off. Then, the local small-scale dynamo operating in the photosphere is the only remaining one, but it cannot produce the same level of activity of the large-scale dynamo, which supports hot loops emitting in the X-rays. In other words, only a low chromospheric emission close to the basal value is expected to remain in this case.

\begin{acknowledgements}  
The authors would like to thank an anonymous Referee for a careful reading of the manuscript and valuable comments that helped to improve their work. They are also grateful to the Editors of this special issue on tides for their kind invitation to contribute. 
S. M. acknowledges funding by the European Research Council through ERC grant SPIRE 647383. This work was also supported by the ANR Blanc TOUPIES SIMI5-6 020 01, the Programme National de Plan\'etologie (CNRS/INSU) and PLATO CNES grant at Service d'Astrophysique (CEA-Saclay). The authors gratefully acknowledge use of the EZ-web stellar evolution resources. Exoplanet studies at  INAF-Osservatorio Astrofisico di Catania have been funded also through the {\it Progetti Premiali} of the Italian {\it Ministero dell'Istruzione, Universit{\`a} e Ricerca}. 
\end{acknowledgements}

\bibliographystyle{spbasic}

\bibliography{LanzaMathis_final}

\end{document}